\documentclass[aps,prd,superscriptaddress,nofootinbib,amsmath,amsfonts,preprintnumbers,groupedaddress,showpacs,10pt,english]{revtex4-1}
\usepackage{amsmath}
\usepackage{amssymb}
\usepackage{babel}
\usepackage{wrapfig}
\usepackage{cancel}

\usepackage{relsize,exscale}
\makeatletter

\usepackage{array,multirow,graphicx}
\usepackage{dcolumn}
\usepackage{newlfont}
\usepackage{bm}
\usepackage[colorlinks,citecolor=blue,urlcolor=blue,linkcolor=blue]{hyperref}
\usepackage[figtopcap]{subfigure}
\usepackage{color}
\usepackage{verbatim}

\def\be{\begin{align}}
\def\ee{\end{align}}
\def\bea{\begin{eqnarray}}
\def\eea{\end{eqnarray}}
\def\bal{\begin{align}}
\def\eal{\end{align}}

\usepackage{scalerel}
\usepackage{tikz}
\usetikzlibrary{svg.path}
\definecolor{orcidlogocol}{HTML}{A6CE39}
\tikzset{
 orcidlogo/.pic={
 \fill[orcidlogocol] svg{M256,128c0,70.7-57.3,128-128,128C57.3,256,0,198.7,0,128C0,57.3,57.3,0,128,0C198.7,0,256,57.3,256,128z};
 \fill[white] svg{M86.3,186.2H70.9V79.1h15.4v48.4V186.2z}
 svg{M108.9,79.1h41.6c39.6,0,57,28.3,57,53.6c0,27.5-21.5,53.6-56.8,53.6h-41.8V79.1z M124.3,172.4h24.5c34.9,0,42.9-26.5,42.9-39.7c0-21.5-13.7-39.7-43.7-39.7h-23.7V172.4z}
 svg{M88.7,56.8c0,5.5-4.5,10.1-10.1,10.1c-5.6,0-10.1-4.6-10.1-10.1c0-5.6,4.5-10.1,10.1-10.1C84.2,46.7,88.7,51.3,88.7,56.8z};}}
\newcommand\orcid[1]{\href{https://orcid.org/#1}{\mbox{\scalerel*{
\begin{tikzpicture}[yscale=-1,transform shape]
\pic{orcidlogo};
\end{tikzpicture}
}{|}}}}
\graphicspath{{./}{Figs/}}
\begin{document}
\date{\today}
\title{Black Hole Topologies and Geodesic Structures in Symmetric Teleparallel  f(Q)  Gravity}

\author{G.G.L. Nashed$^{1,2}$}\email{nashed@bue.edu.eg}
\author{A.~Eid$^{3}$}\email{amaid@imamu.edu.sa}
\affiliation{$^{1}$Centre for Theoretical Physics, The British University, P.O. Box 43, El Sherouk City, Cairo 11837, Egypt\\$^{2}$Center for Space Research, North-West University, Potchefstroom 2520, South Africa\\ $^{3}$Department of Physics, College of Science, Imam Mohammad Ibn Saud Islamic University (IMSIU), Riyadh, Kingdom of Saudi Arabia}

\begin{abstract}
Black hole solutions are studied here within the symmetric teleparallel formulation of gravity, employing the $f(Q)$ model in which the gravitational dynamics are governed by the non-metricity scalar $Q$. We focus on static, circularly symmetric spacetimes in $(2+1)$-dimensions, analyzing both charged and uncharged cases. By adopting a power-law form for $f(Q)$, we derive exact black hole solutions and explore their thermodynamic and geometric properties. Curvature and non-metricity scalars reveal central singularities stronger than those in general relativity. {  we find that the horizon radii increase with the charge parameter while higher values of the non-metricity coefficient, $c_{4}$, or the cosmological constant $\Lambda$ tend to merge or eliminate horizons, reducing their total number and altering the near-origin structure of the spacetime.}  We perform a detailed topological analysis based on the Euler characteristic and examine the geodesic completeness of the spacetime. Our findings show that, depending on the presence of electric charge, the singularity may or may not be reachable by geodesics. The thermodynamic stability is confirmed via temperature, entropy, and heat capacity calculations. This study highlights the rich structure of $f(Q)$ gravity in lower-dimensional settings and offers new insights into the nature of singularities and black hole topologies in modified gravity theories.
\end{abstract}

\maketitle

\section{Introduction}\label{S1}
Einstein's theory of General Relativity (GR) describes gravity using Riemannian geometry. A key aspect of this theory is the choice of the ``Levi-Civita connection'', which ensures that curvature is the only non-zero geometrical quantity. In this standard formulation, both torsion and non-metricity are assumed to be zero. However, other choices of the connection are possible,thus providing different representations that are nevertheless equivalent at the level of gravitational dynamics \cite{BeltranJimenez:2019esp,Harada:2020ikm}. By relaxing these constraints, we can develop theories of gravity where curvature, torsion, and non-metricity can all exist simultaneously. For example, if curvature and non-metricity are zero but torsion remains, we arrive at the Teleparallel Equivalent of General Relativity (TEGR) \cite{Aldrovandi:2013wha,Maluf:2013gaa}. Conversely, if we eliminate torsion but retain non-metricity, we obtain the Symmetric Teleparallel Formulation of General Relativity (STGR) \cite{Nester:1998mp,Adak:2005cd,Adak:2013vwa,Mol:2014ooa,Jarv:2018bgs}.

Despite its success, Einstein's  GR  faces major challenges: explaining the Universe's dark content and quantizing gravity. The search for explanations of the dark sector has encouraged the exploration of modified gravitational models. One common strategy is to extend the Einstein--Hilbert action by replacing its geometric scalar with a more general functional form. In this context, $f(R)$ gravity modifies the action by promoting the Ricci scalar $R$ to an arbitrary function $f(R)$ (see the reviews in \cite{DeFelice:2010aj,Sotiriou:2008rp,Capozziello:2011et,Nojiri:2010wj}).

An alternative route is provided by teleparallel gravity, where gravitational interactions are described through torsion rather than curvature. The teleparallel equivalent of General Relativity (TEGR) forms the basis of this formulation and allows the construction of extensions such as $f(T)$ gravity \cite{Bengochea:2008gz,yousaf2025dark,Nashed:2019yto,alshammari2025mass,Linder:2010py,Cai:2015emx,Nojiri:2017ncd,Lu:2019hra}.

Following a similar approach, $f(Q)$ gravity can be formulated by including an arbitrary function of the non-metricity Q in the gravitational Lagrangian \cite{BeltranJimenez:2018vdo}. An attractive property of $f(Q)$ gravity is that, similarly to $f(T)$ gravity, its field equations remain of second order, while $f(R)$ theories generally lead to fourth-order dynamics. Because of this feature, $f(Q)$ models have been widely investigated in cosmology. Studies include dynamical analyses \cite{Lu:2019hra,yousaf2025electromagnetic}, observational reconstructions using $f(z)$ techniques \cite{Lazkoz:2019sjl}, and cosmographic approaches \cite{Mandal:2020buf}.

At the background level, the predictions of $f(Q)$ cosmology often mimic those of $f(T)$ gravity \cite{BeltranJimenez:2019tme,Nashed:2010ocg}. To distinguish between these scenarios, several aspects have been explored, such as Noether symmetries \cite{BeltranJimenez:2019tme}, energy conditions \cite{alshammari2025stability,ElHanafy:2014efn,Mandal:2020lyq}, and cosmological perturbations \cite{BeltranJimenez:2019tme,Barros:2020bgg,Shirafuji:1996im}. In addition, $f(Q)$ gravity has been applied to bouncing cosmologies \cite{Bajardi:2020fxh} and embedded within broader modified gravity constructions \cite{Jarv:2018bgs,Gakis:2019rdd,Lazkoz:2019sjl,yousaf2024quasi,yousaf2025viscous,Xu:2019sbp,yousaf2025construction,Xu:2020yeg,DAmbrosio:2020nev}.

Given its ability to reproduce the observed late-time acceleration, it is important to test $f(Q)$ gravity beyond cosmological backgrounds. Astrophysical observations provide a promising avenue, particularly through gravitational-wave detections \cite{LIGOScientific:2016aoc} and the Event Horizon Telescope image of the Galactic center black hole \cite{EventHorizonTelescope:2019dse}.

The Banados-Teitelboim-Zanelli (BTZ) black hole, a solution in $(2+1)$-dimensions with a negative cosmological constant, provides valuable insights into black hole physics, quantum gravity, and its connections to string theory \cite{Banados:1992wn,Banados:1992gq}. This black hole serves as a prototype for the general AdS/CFT correspondence \cite{Maldacena:1997re, Witten:1998qj} and helps us understand the dynamics of spacetime at the event horizon. Studying BTZ black holes offers a simplified framework for investigating problems like information loss and the endpoint of quantum evaporation  challenges that are more readily addressed in low-dimensional models. Furthermore, BTZ black holes are linked to asymptotically flat black strings via T- and U-duality \cite{Horowitz:1993jc,Horne:1991gn}. This study utilizes the general form of $f(Q)$ to investigate charged, spherically symmetric $(2+1)$-dimensions, incorporating constraints such as $g_{tt}\neq 1/g_{rr}$. We will analyze the thermal stability of this model and its singularities. The application of generalized free energy combined with Duan's $\phi$-mapping topological current theory reveals the existence of two distinct topological classes in the BTZ spacetime.  { This work provides the first explicit a charged black hole solution in $(2+1)$-dimensional spacetime formulated in the framework of $f(Q)$ gravity without imposing the constraint $g_{tt}=1/g_{rr}$. Unlike previous studies that focused on uncharged or constrained forms of the metric, our solution reveals a richer causal and topological structure governed by non-metricity corrections. In addition, we explore thermodynamic and topological stability, extending the analysis beyond previous $f(T)$ or $f(Q)$ black-hole models.}

This work examines the existence of charged, spherically symmetric black hole solutions in $(2+1)$-dimensional spacetime within the framework of $f(Q)$ gravity, without introducing prior restrictions on the functional form of $f(Q)$. By adopting the coincident gauge condition, we determine an explicit form of the $f(Q)$ function in $(2+1)$ dimensions and obtain a black hole solution that differs from the well-known BTZ solution of General Relativity (GR). Interestingly, the obtained analytical solution admits de Sitter-type spacetimes even though the field equations do not explicitly include a cosmological constant. The properties and potential advantages of this solution are then examined. This study is organized as: In Sec.~\ref{S1}, we  present the fundamentals of symmetric teleparallel geometry and outline the main features of $f(Q)$ gravity. In Sec.~\ref{S2}, we consider a $(2+1)$-dimensional spherically symmetric spacetime and derive the corresponding charged field equations in $f(Q)$ gravity. By solving the resulting system of differential equations without imposing constraints on the $f(Q)$ function, we show that the resulting geometry differs from the charged BTZ black hole due to higher-order corrections associated with non-metricity. Nevertheless, the solution reduces to the GR limit under suitable conditions.
In Sec.~\ref{S4}, the physical characteristics of the obtained solution are analyzed and its differences from the GR counterpart are discussed. The thermodynamic behavior of the solution is investigated in Sec.~\ref{S41}, where its thermodynamic stability is demonstrated. In Sec.~\ref{S5}, we construct the parameter space corresponding to the solution derived in Sec.~\ref{S2} by employing a generalized free energy approach, and we determine the associated zero points together with their topological charges. In Subsec.~\ref{S661}, the possibility of generating black holes with multiple horizons is explored. Finally, the last section summarizes the main physical results and conclusions of this study.

\section{Maxwell-$f(Q)$ framework}
\label{S1}

 The geometry of Weyl, extends the structure of Riemannian geometry, which provides the geometric foundation of General Relativity (GR). Within Weyl geometry, the parallel transport of a vector around a closed curve leads to variations not only in its orientation but also in its length. As a result, the covariant derivative of the metric tensor does not vanish in Weyl's formulation. This property is characterized by a geometric quantity known as non-metricity, denoted by $Q$. 
 
Accordingly, when the metric tensor $g_{\mu\nu}$ is covariantly differentiated with respect to a general affine connection $\overline{\Gamma}^{\sigma}_{~\mu\gamma}$, one obtains the non-metricity tensor $Q_{\gamma\mu\nu}$, which is defined as  \cite{BeltranJimenez:2017tkd,BeltranJimenez:2019tme, BeltranJimenez:2018vdo},%
\begin{equation}
Q_{\gamma \mu \nu }=-\nabla _{\gamma }g_{\mu \nu }=-\frac{\partial g_{\mu
\nu }}{\partial x^{\gamma }}+g_{\nu \sigma }\overline{\Gamma }_{~\mu \gamma
}^{\sigma }+g_{\sigma \mu }\overline{\Gamma }_{~\nu \gamma }^{\sigma }.
\label{2f}
\end{equation}
Consequently, the general affine connection in Weyl geometry can be represented through a Weyl connection that splits into two independent terms, written as,
\begin{equation}\label{111}
\overline{\Gamma }_{\ \mu \nu }^{\gamma }=\left\{ {}^{\, \alpha}_{\mu \nu} \right\}+L_{\ \mu \nu }^{\gamma }.
\end{equation}%
As shown in Eq. (\ref{111}), the metric $g_{\mu \nu }$  determines the Levi-Civita connection appearing in the first term, which takes the form
\begin{equation}
\left\{ {}^{\, \alpha}_{\mu \nu} \right\}\equiv\frac{1}{2}g^{\gamma \sigma }\left( \frac{%
\partial g_{\sigma \nu }}{\partial x^{\mu }}+\frac{\partial g_{\sigma \mu }}{%
\partial x^{\nu }}-\frac{\partial g_{\mu \nu }}{\partial x^{\sigma }}\right).
\end{equation}

The contribution arising from non-metricity in Eq. (\ref{111}) is contained in the second term, which defines the disformation tensor and may be written as,
\begin{equation}\label{dis}
L_{~\mu \nu }^{\gamma }\equiv\frac{1}{2}g^{\gamma \sigma }\left( Q_{\nu \mu
\sigma }+Q_{\mu \nu \sigma }-Q_{\gamma \mu \nu }\right) =L_{~\nu \mu
}^{\gamma }.
\end{equation}%
By taking appropriate contractions of the non-metricity tensor, one obtains the scalar quantity associated with non-metricity from the disformation tensor, written as,
\begin{equation}\label{QM}
Q\equiv -g^{\mu \nu }\left( L_{\ \ \beta \mu }^{\alpha }L_{\ \ \nu \alpha
}^{\beta }-L_{\ \ \beta \alpha }^{\alpha }L_{\ \ \mu \nu }^{\beta }\right) .
\end{equation}

The inclusion of spacetime torsion extends Weyl geometry into the Weyl-Cartan framework, where torsion plays a fundamental role. In such a geometry, the general affine connection can be written as two distinct terms,
\begin{equation}\label{222}
\overline{\Gamma }_{\ \mu \nu }^{\gamma }=\left\{ {}^{\, \alpha}_{\mu \nu} \right\}+K_{\ \mu \nu }^{\gamma }.
\end{equation}%
As indicated in Eq. (\ref{222}), the term appearing on the right-hand side denotes the contortion tensor derived from the torsion tensor $T_{\ \mu \nu }^{\gamma }$. Here, the torsion tensor is defined through the antisymmetric part of the connection $2\overline{\Gamma }_{~[\mu \nu ]}^{\gamma }$,  and can be written as, 
\begin{equation}
K_{\ \mu \nu }^{\gamma }\equiv \frac{1}{2}g^{\gamma \sigma }\left( T_{\mu
\sigma \nu }+T_{\nu \sigma \mu }+T_{\sigma \mu \nu }\right) .
\end{equation}

A relation can be established between the curvature tensors $R_{\sigma \mu \nu }^{\rho }$ and $\mathring{R}_{\sigma \mu \nu }^{\rho }$  which are associated with the affine connections $\overline{\Gamma }_{\ \mu \nu }^{\gamma }$ and $\Gamma _{\ {\mu \nu} }^{\gamma }$ and it is expressed as
\begin{equation}
R_{\sigma \mu \nu }^{\rho }=\mathring{R}_{\sigma \mu \nu }^{\rho }+\mathring{%
\nabla}_{\mu }L_{\nu \sigma }^{\rho }-\mathring{\nabla}_{\nu }L_{\mu \sigma
}^{\rho }+L_{\mu \lambda }^{\rho }L_{\nu \sigma }^{\lambda }-L_{\nu \lambda
}^{\rho }L_{\mu \sigma }^{\lambda },
\end{equation}%
\begin{equation}
R_{\sigma \nu }=\mathring{R}_{\sigma \nu }+\frac{1}{2}\mathring{\nabla}_{\nu
}Q_{\sigma }+\mathring{\nabla}_{\rho }L_{\nu \sigma }^{\rho }-\frac{1}{2}%
Q_{\lambda }L_{\nu \sigma }^{\lambda }-L_{\sigma \lambda }^{\rho }L_{\rho
\sigma }^{\lambda },  \label{2g}
\end{equation}%
and the relationship regarding Ricci scalar  has the from:%
\begin{equation}
R=\mathring{R}+\mathring{\nabla}_{\lambda }Q^{\lambda }-\mathring{\nabla}%
_{\lambda }\tilde{Q}^{\lambda }-\frac{1}{4}Q_{\lambda }Q^{\lambda }+\frac{1}{%
2}Q_{\lambda }\tilde{Q}^{\lambda }-L_{\rho \nu \lambda }L^{\lambda \rho \nu
}.  \label{2h}
\end{equation}%
Here $\mathring{\nabla}$ denotes the covariant derivative operator constructed from the Levi-Civita connection $\Gamma _{\ \mu \nu }^{\gamma }$.  In the framework of  STEGR, the general affine connection is subject to the conditions of vanishing curvature and torsion. In particular, the Riemann tensor $R_{\sigma \mu \nu }^{\rho }\left( \overline{\Gamma }\right)$ must identically vanish in order to satisfy the curvature-free requirement. When the Riemann tensor is zero, the parallel transport defined by the covariant derivative $\nabla$ becomes independent of the chosen path.

Within STEGR, the torsion tensor is also required to vanish, $T_{\ \mu \nu }^{\gamma }=0$, implying that the affine connection is torsion-free in addition to being curvature-free. Consequently, gravitational interactions in this theory are entirely described by non-metricity. The absence of torsion further ensures that the lower indices of the general affine connection are symmetric. 

 We conclude this section by recalling that there exists a special gauge choice in which STEGR,
where the connection can be arbitrarily chosen, can be cast in a particularly simple form: The so called coincident
gauge \cite{BeltranJimenez:2017tkd}. In this gauge, the connection is trivial, i.e. $\overline{\Gamma }_{\ \mu \nu }^{\gamma }=0$. It is obtained by observing that the first postulate of
STEGR, the vanishing of curvature, implies that the connection must have the form

\begin{equation}
\overline{\Gamma }_{\ \mu \nu }^{\gamma }=(\Lambda^{-1})^\gamma{}_\rho \,\partial_\mu \Lambda^\rho{}_\nu
\end{equation}
where $\Lambda^\rho{}_\nu \in2 GL(4;R)$. The requirement of vanishing torsion further restricts the matrix $\Lambda^\rho{}_\nu$ to have the form $\Lambda^\rho{}_\nu=\partial_\nu \xi^\rho$, for arbitrary $\xi^\rho$, and the connection consequently becomes:
\begin{equation}
\overline{\Gamma }_{\ \mu \nu }^{\gamma }=\frac
{\partial x^{\gamma }}{\partial y^{\beta }}\partial _{\mu }\partial _{\nu} \xi^\beta.\end{equation}

Hence, the connection can be set globally to zero by the affine gauge choice $\xi^\rho=M^\rho{}_\mu x^\mu+ \xi^\rho{}_0$ where $M^\rho{}_\mu$ is a
non-degenerate matrix with constant entries and $\xi^\rho{}_0$ is a constant vector \cite{BeltranJimenez:2018vdo}.

Based on the non-metricity scalar, the action corresponding to $f(Q)$ gravity can be expressed as follows \cite{BeltranJimenez:2017tkd}:
\begin{equation}\label{action1}
    S=-\frac{1}{2\kappa}\int_\mathcal{M}   f({Q}) \sqrt{-g} d^4x +\int \sqrt{-g} {\cal L}_{ em}~d^{4}x \,.
\end{equation}
The spacetime manifold is denoted by  $\mathcal M$, while  $g_{\mu\nu}$, represents the covariant metric tensor and  $g$, corresponds to its determinant. The function respectively  $f({Q})$ characterizes a general functional dependence on the non-metricity scalar
$Q$. In relativistic units where $c=G=1$, the constant $\kappa$ is defined as  $\kappa= 8 \pi$ 

In Eq. (\ref{action1}), the electromagnetic sector is described by the Maxwell Lagrangian    ${\cal L}_{
em}=-\frac{1}{2}{ F}\wedge ^{\star}{F}$  where the field strength 2-form is given by $F = d\varphi$ and  $\varphi=\varphi_{\mu}dx^\mu$ denotes the electromagnetic potential 1-form \cite{Awad:2017tyz}. 

Similar to
$f(R)$ theory, the $f(Q)$ framework presents modifications to Einstein's General Relativity. In particular, the theory reduces to the  STEGR  when the function takes the simple form
$f(Q)=Q$.
 
 Due to the symmetry of the metric tensor $g_{\mu \nu}$, the non-metricity tensor
$Q_{\gamma \mu \nu }$, admits only two independent traces, which can be obtained from its contractions
\begin{equation}
Q_{\gamma }\equiv Q_{\gamma \ \ \ \mu }^{\ \ \mu },\ \ \ \ \tilde{Q}_{\gamma
}\equiv Q_{\ \ \gamma \mu }^{\mu }.  \label{2l}
\end{equation}

For completeness, we introduce the superpotential (conjugate) corresponding to the non-metricity tensor, which is written as,

\begin{equation}
\hspace{-0.5cm}P_{\ \ \mu \nu }^{\gamma }\equiv \frac{1}{4}\bigg[-Q_{\ \ \mu
\nu }^{\gamma }+2Q_{\left( \mu \ \ \ \nu \right) }^{\ \ \ \gamma }+Q^{\gamma
}g_{\mu \nu }-\widetilde{Q}^{\gamma }g_{\mu \nu }-\delta _{\ \ (\mu
}^{\gamma }Q_{\nu )}\bigg]=-\frac{1}{2}L_{\ \ \mu \nu }^{\gamma }+\frac{1}{4}%
\left( Q^{\gamma }-\widetilde{Q}^{\gamma }\right) g_{\mu \nu }-\frac{1}{4}%
\delta _{\ \ (\mu }^{\gamma }Q_{\nu )}.
\end{equation}

From the above relations, the expression for the non-metricity scalar is obtained as,
\begin{equation}\label{Qs}
Q=-Q_{\gamma \mu \nu }P^{\gamma \mu \nu }=-\frac{1}{4}\big(-Q^{\gamma \nu
\rho }Q_{\gamma \nu \rho }+2Q^{\gamma \nu \rho }Q_{\rho \gamma \nu
}-2Q^{\rho }\tilde{Q}_{\rho }+Q^{\rho }Q_{\rho }\big).
\end{equation}
The equations of motion are derived by independently making variation of Eq.~(\ref{action1}) w.r.t. the metric as well as the matter sector, resulting in \cite{Heisenberg:2023lru}:
\begin{align}\label{1st EOM}
&\zeta_{\mu \nu}=\frac{2}{\sqrt{-g}} \nabla_\alpha \left( \sqrt{-g} f_{Q} P^\alpha_{\;\; \mu \nu }\right) + \frac{1}{2} g_{\mu \nu} f + f_{Q} \left( P_{\mu \alpha \beta } {Q}^{\;\;\alpha \beta}_\nu
- 2P_{\alpha \beta \mu} {Q}^{\;\;\alpha \beta}_\nu\right)+\kappa^2\frac{1}{2}\kappa{{{\cal
T}^{{}^{{}^{^{}{\!\!\!\!\scriptstyle{em}}}}}}}_{\mu \nu}\nonumber\\
&\partial_\nu \left( \sqrt{-g} F^{\mu \nu} \right)=0\;,
\end{align}
 Taking the functional derivative of the action in Eq.~(\ref{action1}) with respect to the connection gives the following equation,
\begin{equation}\label{2nd EOM}
{ \nabla^\mu \nabla^\nu \left(\sqrt{-g} f_{Q} P^\alpha_{\;\; \mu \nu }\right)=0}\,.
\end{equation}
In this study, ${\mathcal{T}_\mu^\nu{}^{^\text{em}}}$ denotes the tensor characterizing the electromagnetic field's energy-momentum, which is figured as:
\begin{align}\label{Max}
{{{\cal
T}^{{}^{{}^{^{}{\!\!\!\!\scriptstyle{em}}}}}}}^\nu_\mu=F_{\mu \alpha}F^{\nu \alpha}-\frac{1}{4} \delta_\mu{}^\nu F_{\alpha \beta}F^{\alpha \beta}.\end{align}
In Eq.~(\ref{1st EOM}), the function $f$ corresponds to $f(Q)$, while $f_{Q}$ is the first derivative of the function $f$ w.r.t. the non-metricity scalar $Q$, i.e., $f_{Q}=\frac{df}{dQ}$. It is important to note that the matter Lagrangian density is assumed to be independent of the affine connection. Therefore, its variation with respect to the connection vanishes, and as a consequence the hypermomentum tensor does not appear in the resulting field equations.

Furthermore, it is well known that the predictions of General Relativity can be recovered within the  STEGR  by choosing $f(Q)=Q$. In this particular case, the gravitational Lagrangian density reduces to
\begin{equation}
\mathcal{L} = -\frac{Q}{2\kappa^{2}} + \mathcal{L}_{em}.
\end{equation}

\section{(2+1)-dimension Charged Spherically symmetric spacetime}\label{S2}
The $(2+1)$-dimensional spherically symmetric spacetime metric can be expressed as\footnote{When the metric functions are not equal i.e.,  $k(r)\neq k_1(r)$, the structure of the spacetime becomes significantly altered. This asymmetry impacts the relationship between light and matter, leading to distortions of causal cones and horizons, and potentially revealing a richer and more complex causal network than a simple black hole geometry.} \cite{Nashed:2021pkc,Nashed:2021gkp},
\begin{equation}\label{1}
ds^{2}=-k(r)dt^{2}+\frac{dr^{2}}{k_1(r)}+r^{2}d\xi^{2}\,,
\end{equation}
where $k\equiv k(r)$ and $k_1\equiv k_1(r)$ are unknowns. Employing Eqs.~(\ref{1}) and (\ref{QM}), the non-metricity scalar is determined as,
\begin{align}\label{Qc}
Q=-\frac{k_1k'}{rk}\,.
\end{align}
 In $f(Q)$ gravity, the field equations are expressed as follows,
\begin{align}\label{fe1}
&\zeta_t{}^{t}\equiv 0=\frac {2 f_{Q Q} Q'kk_1  +f_Q k_1' k  +r \left[k   \left( f-Q f_Q \right) \right] -2\varphi'^{2}k_1 r}{2rk }, \nonumber\\
&\zeta_r{}^{r}\equiv0=\frac{k  fr-2k  f_{Q}Q  r-2\varphi'^{2}k_1 r}{2rk  }\,,\nonumber\\
&\zeta_\xi{}^{\xi}\equiv0=\frac {2\,f k^{2}-2\,f_Q\,Q   k^{2 }+f_Qk_1'  k'   k  +2f_Q\,k_1   k''k  -f_Qk_1   k'^{2}  +2f_{QQ} k_1 k' Q' k +4\varphi'^{2}k k_1 }{4  k^{2}}\,,
\end{align}
where the prime symbol refers to differentiation with respect to the radial coordinate $r$.
To construct the general solution of the above differential system, we utilize the chain rule as\footnote{Because of the imposed spherical symmetry, the functional dependence $f(Q(r))$ can be expressed simply as $f(r)$.}
\begin{eqnarray}\label{dfg}
&&f(Q)=f(r),\nonumber\\
& & f_Q=\frac{df(Q)}{dQ}=\frac{df(r)}{dr}\frac{d r}{dQ}=-\frac { f' {r}^{2} k^{2}}{ k''krk_1 - k'^{2} k_1  r+k' k'_1rk -k_1 k' k },\nonumber\\
& & f_{QQ}=\frac{df_Q}{dQ}=\frac{d}{dr}\Bigg(\frac{df(r)}{dr}\frac{d r}{dQ}\Bigg)=-\frac {1}{ \left[  k''k_1 rk + k' \left\{k  \left( k'_1 r -k_1 \right) - k' k_1 r \right\}  \right] ^{3}}\left[k^{3} k rk''\left\{  f' {r}^{2} k^{2}k''' k_1 - \left\{ k rf'' k_1\right.\right.\right.\nonumber\\
 &&\left.\left.\left.  -2\, f'  \left(k  \left[  k'_1 r-k_1 \right]  \right) -\frac{3}2\, k' k_1 r  \right\}  + \left( k \left\{  k' k_1 r-k  \left( k'_1 r-k_1 \right)  \right\} rf'' + \left( {r}^{2} k^{2}k''_1 +2\,{r}^{2}k_1 k'^{2}-2\,rk  \left(  k'_1 r-k_1  \right) k'\right.\right.\right.\right.\nonumber\\
 &&\left.\left.\left.\left.  -2\, k^{2} \left( k'_1r -k_1\right)  \right) f' \right) k'  \right\} {r}^{3}\right] ,\nonumber\\
& &
\end{eqnarray}
where $f'=\frac{df(r)}{dr}$ and  $f''=\frac{d^2f(r)}{dr^2}$.
The system of the differential equation \eqref{fe1} is a closed system four non-linear differential equations in four unknown, $k$, $k_1$, $f$ and $\varphi$. Thus, the system admits an exact solution of the form, 
\begin{align}\label{sol}
&k(r)=c_1+c_5\int \! \frac{\left( {r}^{2}c_3-2{c_2}^{2}+rc_4\right) ^{2}\frac{e^{-4c_4{\tanh^{-1}} \left( {\frac {2
c_3r+c_4}{\sqrt {8{c_2}^{2}c_3+{c_4}^
{2}}}} \right)}}{\sqrt {8{c_2}^{2}c_3+{c_4
}^{2}}}}{{r}^{3}}{dr}, \qquad k_1(r)=\frac{r^4\frac{e^{8c_4{\tanh^{-1}} \left( {\frac {2
c_3r+c_4}{\sqrt {8{c_2}^{2}c_3+{c_4}^
{2}}}} \right)}}{\sqrt {8{c_2}^{2}c_3+{c_4
}^{2}}}}{\left( {r}^{2}c_3-2{c_2}^{2}+rc_4\right) ^{2}}\times k(r)\,,\nonumber\\
&f=c_3+\frac{c_4}{r}, \qquad \varphi=c_6+c_2\int \! \frac{\left(2{c_2}^{2}-{r}^{2}c_3-rc_4\right)\frac{e^{-4c_4{tanh^{-1}} \left( {\frac {2
c_3r+c_4}{\sqrt {8{c_2}^{2}c_3+{c_4}^
{2}}}} \right)}}{\sqrt {8{c_2}^{2}c_3+{c_4
}^{2}}}}{{r}^{3}}{dr}\,,
\end{align}
where $c_i, \, i=1\cdots 6$ are  constants of integration.
{
We now analyze and  discuss the following results:\\
\begin{itemize}
\item If the constant $c_2=0$ then the charge becomes constant, i.e., $\varphi=c_6$. In that case  $k$ and $k_1$ take the following form,
\begin{align}\label{sol1}
&k(r)=c_1+c_5\int\left( {r}^{2}c_3+rc_4\right)^{2}\frac{e^{-4c_4{\tanh^{-1}} \left( {\frac {2
c_3r+c_4}{c_4}} \right)}}{{r}^{3}}{dr}, \,\, k_1(r)=\frac{r^4e^{8c_4{\tanh^{-1}} \left( {\frac {2
c_3r+c_4}{c_4}} \right)}}{\left( {r}^{2}c_3+rc_4\right) ^{2}}\times k(r)\,, \,\, f=c_3+\frac{c_4}{r}, \,\,\varphi=c_6,
\end{align}
which is the non-charged solution with non-vanishing value of $f$. When the constant $c_4=0$ we get  $f=1$ $k(r)=k_1(r)=c_1+c_5r^2$, which is the BTZ of GR provided $c_3=1$. \\
\item When the constant $c_4=0$ we get $f=1$, provided $c_3=1$, and \begin{align}\label{fc}
\varphi=c_6-c_2\ln(r)-\frac{c_2{}^3}{r^2}\,,\quad k=c_1+c_5\left[\frac{r^2}{2}-4c_2^2\ln(r)-\frac{2c_2{}^4}{r^2}\right]\,, \quad k_1=\frac{1}{\left(1-2\frac{c_2^2}{r^2}\right)^2}\times k\,.\end{align}
\end{itemize}
which is different from charged BTZ. The reason for this is the fact that the anstazs $k$ and $k_1$ are not equal.  When $c_2=0$  we get $k=k_1$ and in that case we get the non-charged BTZ of GR. So in general, we can conclude that when $f(Q)=const$ the charged output solution is not the one of GR because of the un-equality of the metric anstazs.

\medskip
\medskip
\noindent\textbf{Concise parameter--regime map.}
Table~\ref{tab:regimes} summarizes the main limits and the associated geometric behavior.

\begin{table}[h!]
\centering
\caption{Parameter choices and geometric regimes in the solutions of Eqs.~(24)--(27). Here $c_1=-m$, $c_5=\Lambda$.}
\label{tab:regimes}
\begin{tabular}{l l l l l l}
\hline
\textbf{Choice} & \textbf{$f(Q)$} & \textbf{Metric branch} & \textbf{Asymptotics} & &\textbf{Horizon structure}\\
\hline
$c_2=0,\;c_4=0$, &
$1$ (const) &
$k=k_1=c_1+\Lambda r^2$  &
(A)dS by sign of $\Lambda$ &&
BTZ: typically one horizon for $m>0$ \\$c_3=1$& &(BTZ) && & extremal/naked at thresholds\\
$c_2\neq 0,\;c_4=0$, &
$1$ (const) &
\underline{Charged, $k\neq k_1$}  &
(A)dS by sign of $\Lambda$ &&
Branch differs from GR charged BTZ\\$c_3=1$&&(Eq.~(26))&& &$0/1/2$ horizons depending on $(m,c_2,\Lambda)$\\
$c_2=0,\;c_4\neq 0$ &
$c_3+\dfrac{c_4}{r}$ &
Uncharged, $k=k_1$  &
AdS in Sec.~IV  &&
$0/1/2$ horizons depending on \\&&(Eq.~(25))&(our sign; $\Lambda>0$)&&$(m,c_3,c_4,\Lambda)$\\
$c_2\neq 0,\;c_4\neq 0$ &
$c_3+\dfrac{c_4}{r}$ &
Charged, generic  &
AdS in Sec.~IV  &&
Multi-horizon possible (up to three)\\&& $k\neq k_1$ (Eq.~(24))&(our sign; $\Lambda>0$)&&cf.\ Sec.~VI.A/Fig.~4\\
\hline
\end{tabular}
\end{table}
}

\subsection*{Interpretation of the Integration Constants}

The constants \( c_1, c_2, \dots, c_6 \) appearing in the exact solution given by Eq. \eqref{sol} can be assigned the following physical meanings based on their functional roles:

\begin{itemize}
  \item \( \mathbf{c_1} \): This constant appears additively in the temporal component \( g_{tt} \) of the metric function \( k(r) \), and is later identified with \( -m \). Therefore, it plays the role of the black hole's gravitational mass.

  \item \( \mathbf{c_2} \): This constant arises in the expression for the electromagnetic potential \( \phi(r) \). Setting \( c_2 = 0 \) nullifies the electric field, reducing the solution to the uncharged case. Hence, \( c_2 \) is associated with the black hole's electric charge.

  \item \( \mathbf{c_3} \): This constant appears in   \( f(Q) = c_3 + \frac{c_4}{r} \). It affects the asymptotic behavior of the geometry and corresponds to the background gravitational contribution, akin to a cosmological term.

  \item \( \mathbf{c_4} \): This coefficient governs the \( \frac{1}{r} \) correction in \( f(Q) \) and influences the multi-horizon structure. It can be interpreted as a parameter encoding non-metricity corrections or coupling between curvature and electromagnetic effects.

  \item \( \mathbf{c_5} \): This constant is set to \( \Lambda \) in the metric expansion given by Eq. \eqref{fc}, and therefore corresponds to the cosmological constant or the curvature scale of the asymptotically AdS spacetime.

  \item \( \mathbf{c_6} \): Appears as an additive constant in \( \phi(r) \) and represents the gauge freedom in the choice of electromagnetic potential. It does not affect the field strength and hence has no observable physical effect.
\end{itemize}

Understanding these constants is essential for linking the mathematical solution to physical observable and for exploring the parameter space (e.g., horizons, thermodynamic behavior) meaningfully.

{ We now summarizing the six constants into the following table:
\[
\begin{array}{c|l|l}
\text{Constant} & \text{Physical Role} & \text{Interpretation} \\ \hline
c_1 & \text{Mass term} & c_1=-m \\
c_2 & \text{Electric charge} & \text{controls the field strength} \\
c_3 & \text{Background curvature} & \text{acts like a cosmological term} \\
c_4 & \text{Non-metricity correction} & \text{couples geometry and charge} \\
c_5 & \text{Cosmological constant} & \Lambda \\
c_6 & \text{Gauge potential constant} & \text{no physical effect}
\end{array}
\]}
\section{Physical Features of the Black Hole Solution}\label{S4}
In this section, we analyze some physical aspects of the solution derived in the previous section.

\centerline{\underline{\textbf{Metric}}}\vspace{0.2cm}

\begin{figure}
\centering
\subfigure[~The behavior of the temporal component of the metric given by Eq. (\ref{metric})]{\label{fig:R}\includegraphics[scale=0.22]{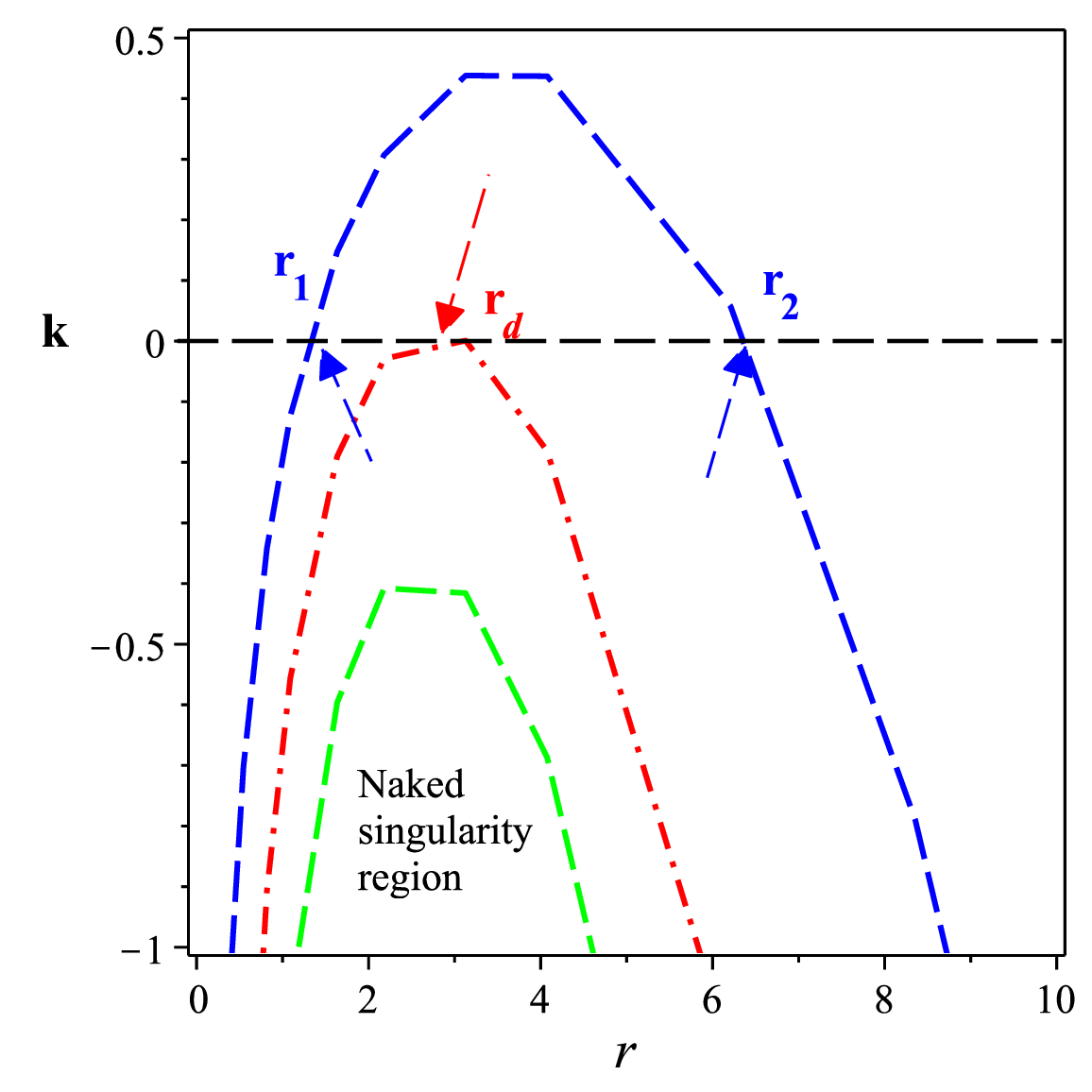}}
\subfigure[~The behavior of the spatial component of Eq. (\ref{metric})]{\label{fig:fr}\includegraphics[scale=0.22]{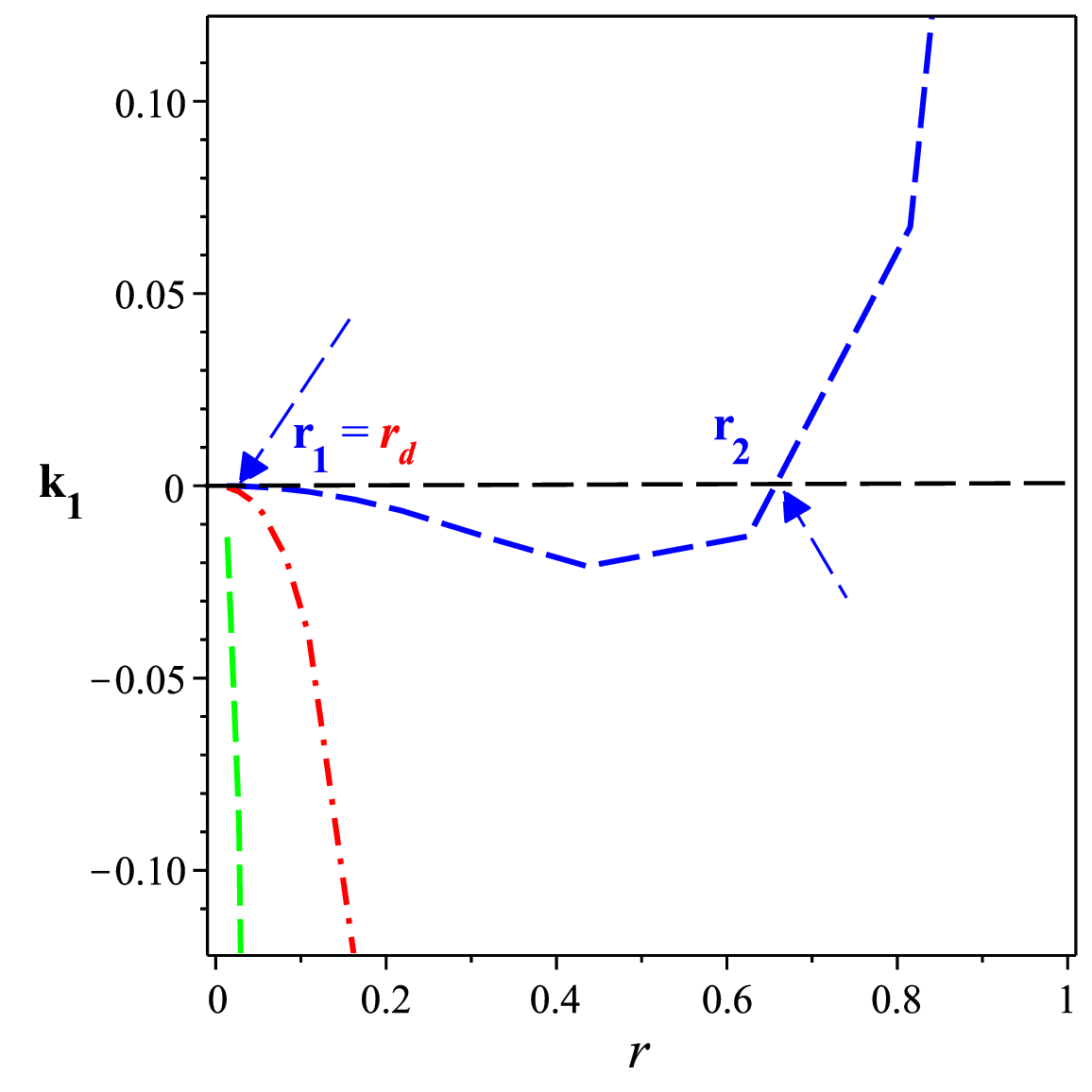}}
\subfigure[~The behavior of the metric ]{\label{fig:fr1}\includegraphics[scale=0.22]{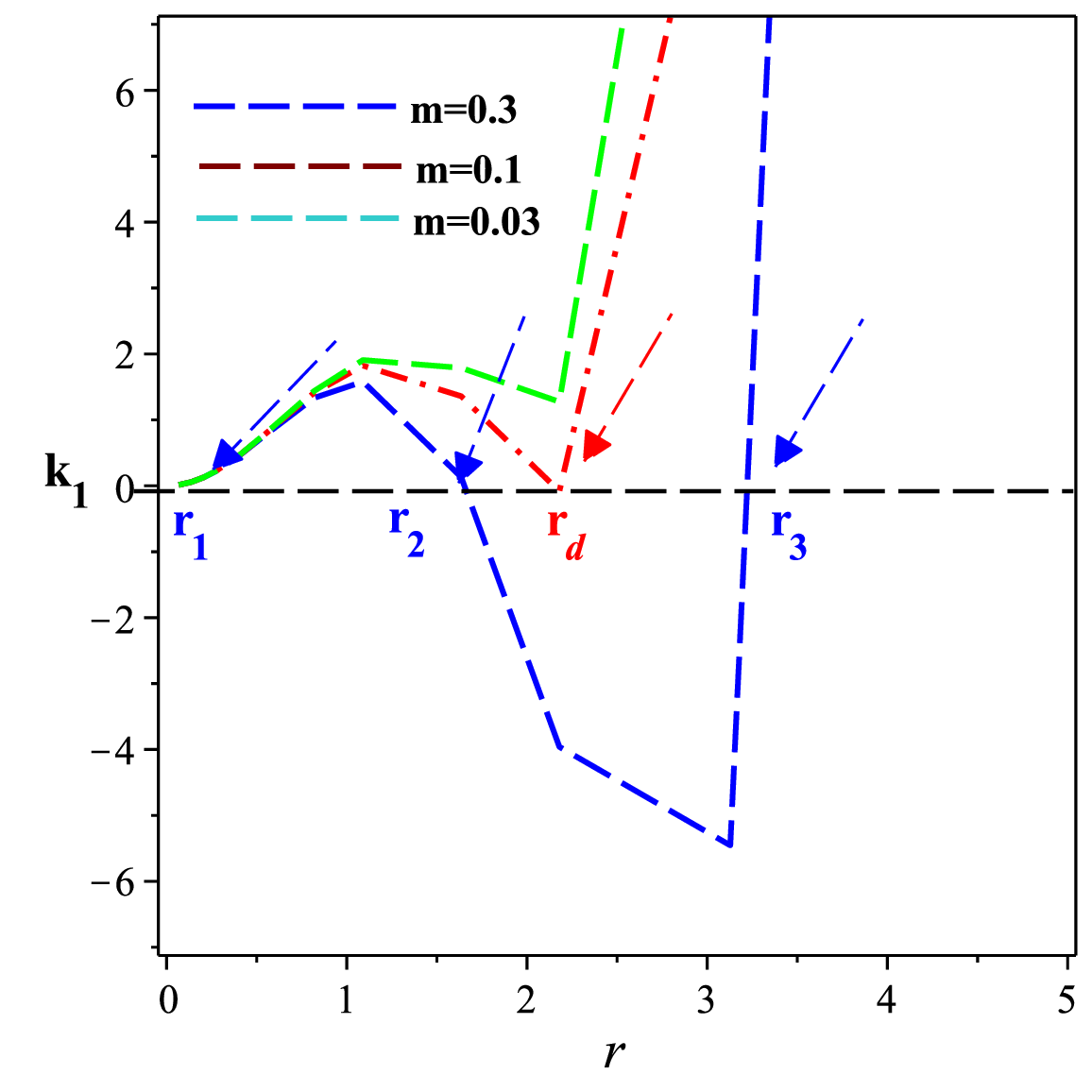}}
\subfigure[~The behavior of the charged field given  by Eq. (\ref{metric})]{\label{fig:c}\includegraphics[scale=0.22]{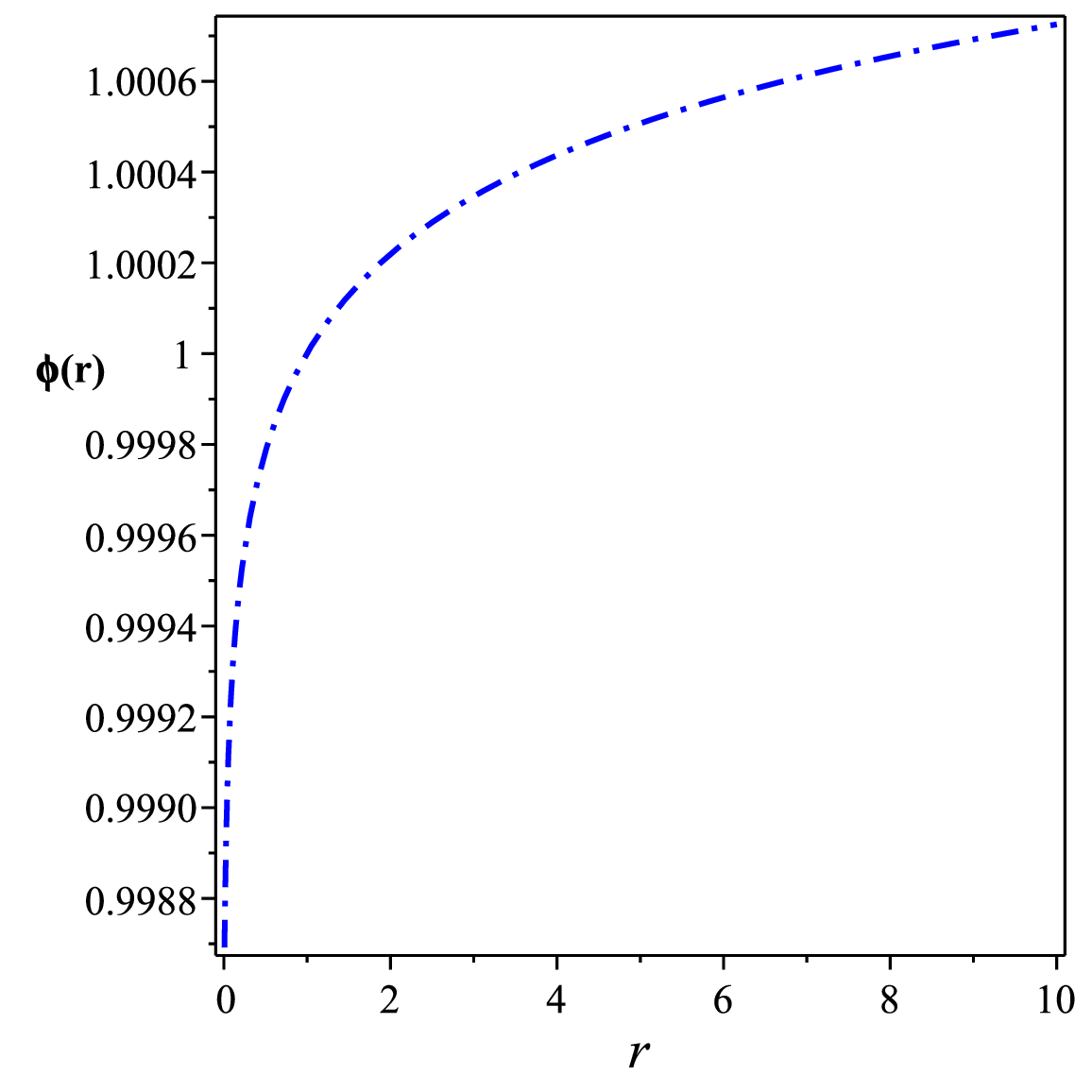}}
\caption[figtopcap]{\small{Systematic graphical representations of the spatial component of  Eq.~(\ref{metric}) are displayed in Fig.~\subref{fig:R}.,  \subref{fig:fr} the temporal component of the metric given by Eq. (\ref{metric}).}}
\label{Fig:1}
\end{figure}

Using Eq.~(\ref{sol}) in Eq.~(\ref{1}), the metric can be written as,

\bea\label{metric}
&& ds^2=-\Biggl[-m+\Lambda \int \! \frac{\left( {r}^{2}c_3-2{c_2}^{2}+rc_4\right) ^{2}{e^{-4\Upsilon}}}{{r}^{3}}{dr}\Biggr]dt^2 +r^2d\xi^2
-\frac{dr^2}{\Biggl\{-m+\Lambda \int \! \frac{\left( {r}^{2}c_3-2{c_2}^{2}+rc_4\right) ^{2}{e^{-4\Upsilon}}}{{r}^{3}}{dr}\Biggr\}\frac{r^4{e^{8\Upsilon}}}{\left( {r}^{2}c_3-2{c_2}^{2}+rc_4\right) ^{2}}}\,,\nonumber\\
&&\approx{e^{-4\Upsilon_1}}\left[\frac{c_5 }{24{c_2}^{4}}\left(12 {c_2}^{4}{c_3}^{2}-26{c_2}^{2}c_3{c_4}^{2}+3{c_4}^{4} \right){r}^{2}+\frac{2c_5}{3{c_2}^{4}} \left(8c_3c_4{c_2}^{4}-3{c_4}^{3}{c_2}^{2} \right) r+\frac{1}{ 24{c_2}^{4}} \left\{ 12c_5 \left( 12{c_2}^{4}{c_4}^{2} -8{c_2}^{6}c_3 \right) \ln( r ) \right.\right.\nonumber\\
&&\left.\left.-24m{c_2 }^{4} \right\} +\frac{8\,{c_2}^{2}c_4c_5}{r}-\frac{2\,{c_2}^{4}c_5}{r^2}\right]dt^2+\left[ \left\{  \left[  \left(\frac{3}2{c_4}^{2}- {c_2}^{2}c_3 \right)  \left( {\frac {21}{4}}{c_4}^{2}+{c_2}^{2}c_3 \right) \ln ( r) -\frac{{c_2}^{4}{c_3}^{2}}4 +{\frac {275}{32}}{c_4}^{4}+{\frac {349}{48}}{c_2}^{2} c_3{c_4}^{2} \right]\right.\right.\nonumber\\
 &&\left.\left.\times c_5{e^{-4\Upsilon}}-\frac{m}4 \left( {\frac {21}{4}} {c_4}^{2}+{c_2}^{2}c_3 \right)  \right\} {e^{8\Upsilon}}{r}^{2}{c_2}^{-8}-{e^{8\Upsilon}} \left\{  \left( 3c_4{c_2}^{2} \left( {c_2}^{ 2}c_3-3/2{c_4}^{2} \right) \ln( r) -\frac{13}2 {c_4}^{3}{c_2}^{2}-{\frac {17}{12}}c_3c_4 {c_2}^{4} \right) c_5{e^{-4\Upsilon}}\right.\right.\nonumber\\
 &&\left.\left.+\frac{3}4 mc_4{c_2}^{2} \right\} r{c_2}^{-8}-\frac{e^{8\Upsilon}}{c_2^8} \left\{  \left( {c_2}^{4} \left( {c_2}^{2}c_3-\frac{3{c_4}^{2}}2 \right) \ln( r) -{\frac {27}{8}}{c_2}^{4}{c_4}^{2}+\frac{1}{2}{c_2}^{6}c_3 \right) c_5{e^{-4\Upsilon}}+\frac{m{c_2}^{4}}4 \right\} \right.\nonumber\\
 &&\left.+\frac{1}2{e^{-4\Upsilon}}c_5{e^{8\Upsilon}}c_4{c_2}^{-2}{r}^{-1}-\frac{1}2{e^{-4\Upsilon}}c_5{e^{8\Upsilon}}{r}^{-2}\right]^{-1}+r^2d\xi^2, \eea
where we have put $c_1=-m$ and $c_5=\Lambda$, $\Upsilon$  and $\Upsilon_1$ are defined as
\begin{align}\label{deff}
\Upsilon= {\frac {c_4 \,\tanh^{-1} \left( {\frac {2c_3r+c_4}{\sqrt {8{c_2}^{2}c_3+{c_4}^{2}}}} \right)}{\sqrt {8{c_2}^{2 }c_3+{c_4}^{2}}}}, \qquad \qquad \Upsilon_1= {\frac {c_4 \,\tanh^{-1} \left( {\frac {c_4}{\sqrt {8{c_2}^{2}c_3+{c_4}^{2}}}} \right)}{\sqrt {8{c_2}^{2 }c_3+{c_4}^{2}}}},\end{align}
 where this constrains should be satisfied $8c_{2}c_{3}+c_{4}^{2}>0$  to ensure real value. The behavior of the metric (\ref{metric}) is shown in Fig.\ref{Fig:1} \subref{fig:R} where we can see that the black hole can have two horizons, the inner one $r_1$ and the outer, $r_2$, one horizon $r_d$ when the two horizons coincide,  and a naked singularity depending on the appropriate choice of the value of $c_3$.
Equation (\ref{metric}) shows clearly that the metric is asymptotically  AdS. By taking the limit $c_4 \rightarrow 0$, we get the charged BTZ  black hole. Notice that although the metric (\ref{metric} )  asymptotically behaves as AdS solutions, which have different $g_{tt}$ and $g^{rr}$ components for the metric, they have coinciding Killing and event horizons. Finally, we show that the charge field has a positive pattern as shown in Fig.\ref{Fig:1} \subref{fig:c}.  From this figure we can show that the finiteness of $\phi(r)$ at the origin is a direct consequence of the
$f(Q)$ gravity modification. Unlike the standard charged BTZ black hole or Reissner-Nordstr\"om solution in GR, where
$\phi(r)~ln(r)$ or diverges as
$r\to 0$, this behavior suggests a non-singular charge distribution. \vspace{0.2cm}\\
 We now investigate the presence of physical singularities by examining the curvature and non-metricity invariants associated with the solution. Special attention is given to the behavior of these invariants near the zeros of the function $k_1(r)$. For this purpose, we evaluate the curvature scalars commonly used in General Relativity along with the non-metricity scalar, which leads to,
\begin{eqnarray*}
 &&{Q}=-\Lambda e^{4\Upsilon}\,,\quad \left({Q}\right)_{r\to \infty}\approx -\frac{\Lambda}{3{{c_3}}^{2}{r}^{3}} \left( 3{c_3}^{2}{r}^{3}+6\,{c_3}\,{r }^{2}{c_4}+3r{{c_4}}^{2}+4{c_4}\,{c_2}^{2} \right) e^{2\Upsilon_1}\,,\nonumber\\
 &&\left({Q}\right)_{r\to 0}\approx -\frac{\Lambda}{12{{c_2}}^{6}} \left( 12{c_2}+12r{c_4}c_2^{4}+9{c_4}^2\,{r }^{2}{c_2}^2+2c_4[c_3c_2^2+3{c_4}^{2}]{r}^{3} \right) e^{2\Upsilon_1}\nonumber\\
&&Q^{\mu \nu \lambda}Q_{\mu \nu \lambda}=2 \left[ 2 \left( {r}^{4}{c_3}^{2}+2{r}^{3}c_3c_4+ \left( -4{c_2}^{2}c_3+2{c_4}^{2} \right) {r} ^{2}+4rc_4{c_2}^{2}+20{c_2}^{4} \right) {r}^{ 4}{\Lambda}^{2} \left( \int \! \left( c_3{r}^{2}+rc_4-2 {c_2}^{2} \right) ^{2}\frac{e^{-4\Upsilon}}{r^3}{dr} \right) ^{2}\right.\nonumber\\
&&\left.+2m{r}^{2}\Lambda \left\{ \Lambda \left( rc_4+4{c_2}^{2} \right)  \left( c_3{r}^{2}+rc_4-2{c_2}^{2} \right) ^{3}{e^{-4\Upsilon}}-2 \left( {r}^{4}{c_3}^{2}+2{r}^{3}c_3c_4+ \left( -4{c_2}^{2}c_3+2{c_4}^{2} \right) {r}^{2} +4rc_4{c_2}^{2}+20{c_2}^{4} \right) {r}^{2}  \right\}\right.\nonumber\\
&&\left. \int \! \left( c_3{r}^{2}+rc_4-2{c_2}^{2} \right) ^{2}\frac{e^{-4\Upsilon}}{r^3}{dr}+{\Lambda}^{2} \left( c_3{r}^ {2}+rc_4-2{c_2}^{2} \right) ^{6}{e^{-8\Upsilon}}+2m \left\{ \Lambda \left( rc_4+4{c_2}^{2} \right)  \left( c_3{r}^{2 }+rc_4-2{c_2}^{2} \right) ^{3}{e^{-4\Upsilon}}\right.\right.\nonumber\\
&&\left.\left.+m \left( {r}^{4}{c_3}^{ 2}+2{r}^{3}c_3c_4+ \left( -4{c_2}^{2}c_3+2 {c_4}^{2} \right) {r}^{2}+4rc_4{c_2}^{2}+20{c_2}^{4} \right) {r}^{2} \right\} {r}^{2} \right] \frac{e^{8\Upsilon}\left\{\int \! \left( c_3 {r}^{2}+rc_4-2{c_2}^{2} \right) ^{2}\frac{e^{-4\Upsilon}}{r^3}{dr}\Lambda-m \right\} ^{-1}}{r^2\left\{ c_3{r}^{2}+rc_4-2{c_2}^{2} \right\} ^{4}} \nonumber\\
&&\left(Q^{\mu \nu \lambda}Q_{\mu \nu \lambda}\right)_{r\to \infty}\approx -2\,{\frac {{\Lambda}^{2}{c_3}^{2}{r}^{2}}{m}}-4\,{\frac {{\Lambda}^{2}c_3\,c_4\,r}{ m}}+2\,{\frac { \left( 4\,{\Lambda}^{2}{c_3}^{3}{c_2}^{2}-{\Lambda}^{2}{c_4}^{2}{c_3}^{2} \right) }{m\,{c_3}^{2}}}+2\,{\frac { \left( 4\,c_4\,{\Lambda}^{2}{c_3}^{2}{c_2}^{2}+2 \,c_4\,\Lambda\,c_3\,+m \right) }{m\,{c_3}^{2}r}}\nonumber\\
&&+2\,{\frac { \left( 8\,+m\,\Lambda\,{c_2}^{2}c_32\,{m}^{2}-2\,m\,\Lambda\,{c_4}^{2}-4\,{\Lambda}^{2}{c_2}^{4}{c_3 }^{2} \right) }{{c_3}^{2}m\,{r}^{2}}}\,,\nonumber\\
&&Q_{\mu}Q^{\mu}=\frac{4r^2e^{8\Upsilon}\left( \Lambda\int \! \left(c_3\,{r}^{2} -2\,{c_2}^{2}+rc_4 \right) ^{2}\frac{e^{-4\Upsilon}}{r^3}{dr}-m \right)  \left( c_3{r}^{2}+2\,rc_4+2\,{c_2}^{2} \right) ^{2}}{ \left(c_3{r}^{2} -2\,{c_2}^{2}+rc_4 \right)^4},
\end{eqnarray*}
\begin{eqnarray*}
&&\left(Q_{\mu}Q^{\mu}\right)_{r\to \infty}\approx {\frac {-4m}{{c_3}^{2}{r}^{2}}}-{\frac
{16\,mc_4}{{c_3}^{3}{r}^{3}}}-{\frac {
4m \left( 12\,{c_2}^{2}c_3+4\,{c_4}^{2} \right) }{{c_3}^{4}{r}^{4}}},\nonumber\\
 &&\left(\tilde{Q}_{\mu}\tilde{Q}^{\mu}\right)_{r\to \infty}\approx 2\,\Lambda\,+\frac{2}{ {c_3}^{2}{\Lambda}{ r}^{2}} \left[ {\Lambda}^{2}{c_3}^{6}\, \left\{ 2\, \left( 2\,{\frac {2\,{c_2}^{2}c_3+{ c_4}^{2}}{{c_3}^{4}}}+{\frac {{c_4}^{2}}{{c_3}^{4} }} \right) {c_3}^{-2}+4\,{\frac {{c_4}^{2}}{{c_3}^{6}}} \right\}-2\,\Lambda\,\, \left( -m-4\,{c_2}^{2}c_3\,\ln  \left( r \right) \Lambda \right)\right.\nonumber\\
  &&\left.-16\,{\Lambda}^{2}{c_4}^{2}+{\frac { \left[ 6\,{\Lambda}^{2}{c_3}^{4}{c_4}^{2}+2\,\Lambda \,{c_3}^{3} \left\{ -4\,\Lambda\,{c_2}^{2}{c_3}^{2} -3\,\Lambda\,c_3\,{c_4}^{2}+\Lambda\, \left( c_3 \, \left( -4\,{c_2}^{2}c_3+{c_4}^{2} \right) +2\,{c_4}^{2}c_3-2\,{c_2}^{2}{c_3}^{2} \right) \right\}  \right] }{{c_3}^{4}}}  \right]
, \nonumber\\
&&\tilde{Q}_{\mu}\tilde{Q}^{\mu}=\left[ -2\,{r}^{2}\Lambda\, \left( 4\,{c_2}^{2}+rc_4 \right) \int \! \left( c_3\,{r}^{2}-2\,{c_2}^{2}+rc_4 \right)^{2}\frac{e^{-4\Upsilon}}{r^3}{dr}+\Lambda\, \left( c_3\,{r}^{2}-2\,{c_2}^{2}+rc_4 \right) ^{3}{e^{-4\Upsilon}}+2\,{r}^{2}m\, \left( 4 \,{c_2}^{2}+rc_4 \right)  \right]^{2}\nonumber\\
&&\times\frac{e^{8\Upsilon}} {\left(c_3{r}^{2} -2\,{c_2}^{2}+rc_4 \right) ^{4}{r}^2\left(\Lambda \int \! \left( -2\,{c_2}^{2}+c_3\,{r}^{ 2}+rc_4 \right) ^{2}\frac{e^{-4\Upsilon}}{r^3}{dr} -m\right) }\,,
\end{eqnarray*}
\begin{eqnarray}
&&R^{\mu \nu \lambda \rho}R_{\mu \nu \lambda \rho}=R^{\mu \nu}R_{\mu \nu}=3 {e^{16\Upsilon}}{\left( c_3{r}^{2}-2 {c_2}^{2}+rc_4 \right) ^6} \left[ \frac{4}3{r}^{4}{\Lambda}^{ 2} \left( 4{c_2}^{2}+rc_4 \right) ^{2} \left( \int \! \left(c_3{r}^{2} -2{c_2}^{2}+rc_4 \right) ^{2}\frac{e^{-4\Upsilon}}{r^3}{dr} \right) ^{2}\right.\nonumber\\
&&\left.-\frac{4}3 \left\{ \Lambda \left( -2{c_2} ^{2}+c_3{r}^{2}+rc_4 \right) ^{3}{e^{-4\Upsilon}}+2m{r}^{2} \left( 4{c_2}^{2}+rc_4 \right)  \right\} {r}^{2}\Lambda \left( 4{c_2}^{2}+rc_4 \right) \int \! \left( c_3{r}^{2}-2 {c_2}^{2}+rc_4 \right) ^{2}\frac{e^{-4\Upsilon}}{r^3}{dr}\right.\nonumber\\
&&\left.+ \left( c_3{r}^{2}-2{c_2}^{2}+rc_4 \right) ^{4}{ \Lambda}^{2} \left( {r}^{4}{c_3}^{2}+\frac{4}3{r}^{3}c_3c_4+ \left( \frac{2}3{c_4}^{2}-\frac{4}3{c_2}^{2}c_3 \right) {r}^{2}-\frac{8}3rc_4{c_2}^{2}+4{c_2}^{4} \right)  {e^{-8\Upsilon}}+\frac{4}3m{r}^{2}\Lambda \left( 4 {c_2}^{2}+rc_4 \right) \right.\nonumber\\
&&\left. \left( c_3{r}^{2}+rc_4 -2{c_2}^{2}\right) ^{3}{e^{-4\Upsilon}}+\frac{4}3{r}^{4}m^{2} \left( 4{c_2}^{2}+rc_4 \right) ^{2} \right]\Rightarrow
\mbox{as} \, \, {r\to \infty} \approx \Lambda^2-\frac{16\Lambda^2c_4}{c_3r}+\frac{52\Lambda^2c_4^2}{3c_3^2r^2}\,,\nonumber\\
&&R= -\frac{3{{e^{8\Upsilon}}}}{ \left( -2{c_2}^{2}+c_3{r}^{2}+rc_4 \right) ^{3}} \left[ \frac{2}3 \left( 4{c_2}^{2}+\Lambda rc_4\right) {r}^{2}\int \! \left( 2{c_2}^{2}-c_3{r}^{2}-rc_4\right) ^{2}\frac{e^{-4\Upsilon}}{r^3}{dr}+\Lambda \left( \frac{2}3rc_4-\frac{2}3{c_2}^{2}+c_3{r}^{2}  \right)\right.\nonumber\\
  &&\left.\left(c_3{r}^{2}-2{c_2}^{2}+rc_4\right) ^{2}{e^{-4\Upsilon}}+\frac{2}3m{r}^{2} \left( 4{c_2}^{2}+rc_4\right)  \right]\Rightarrow
\mbox{as} \, \, {r\to \infty} \approx -3\Lambda-\frac{4\Lambda c_4}{c_3r}-\frac{\Lambda c_4^2}{3c_3^2r^2}. \end{eqnarray}
    Now we are going to discuss that fact that the stronger divergence of curvature invariants compared to general relativity arises because in $f(Q)$ gravity, part of the gravitational degrees of freedom are encoded in the non-metricity sector. While curvature scalars diverge faster, the non-metricity scalar $Q$ remains finite, indicating a redistribution of geometric strength rather than a physical instability. In symmetric teleparallel $f(Q)$ gravity (coincident gauge), the affine connection is trivial and the gravitational dynamics are encoded in the non-metricity sector. The scalar $Q$ aggregates specific contractions of $Q_{\alpha\mu\nu}$ that, for our exact solution, \emph{cancel the leading radial divergences} of the metric derivatives at the origin, leaving $Q$ finite. In contrast, the Levi-Civita curvature scalars (built solely from $g_{\mu\nu}$) respond to the effective source induced by both charge and non-metricity corrections as a stiffer central profile, hence their stronger growth toward $r=0$. Operationally, the modified field equations redistribute part of the ``gravitational load'' from curvature into non-metricity; the particular on-shell combination that defines $Q$ is regular, whereas curvature invariants retain the imprint of the steep central gradients and diverge. This is precisely what we find: $Q$ is nonsingular as $r\to0$, while the curvature sector diverges.
Within our setup, $Q$ is a scalar under diffeomorphisms in the coincident gauge and characterizes the non-metricity sector captured by $f(Q)$. The observed finiteness of $Q$ thus indicates an \emph{improved central regularity for the non-metricity scalar sector}, even though curvature scalars (constructed from the Levi-Civita connection) remain divergent.
\section{Thermodynamics}\label{S41}
{
For the static metric given by Eq.~\eqref{1} with Killing vector $\chi^\mu=\partial_t$, the surface gravity at the outer horizon $r_2$ is
\[
\kappa^2=-\tfrac12(\nabla_\mu\chi_\nu)(\nabla^\mu\chi^\nu)
\quad\Rightarrow\quad
\kappa=\frac12\sqrt{k_{1}(r_2)}\,k'(r_2)
\]
 The Hawking temperature is derived from the surface gravity $\kappa$ at the event horizon $r_2$,
\begin{align} \label{kGR}
T = \frac{\kappa}{2\pi} = \frac{1}{4\pi}\left.\frac{dk(r)}{dr}\right|_{r=r_{2}}.
\end{align}
 In $f(Q)$ gravity the Wald/Noether-charge entropy acquires the standard multiplicative factor $f_Q$ (variation of the Lagrangian density $f(Q)$ with respect to the non-metricity scalar), so in $2{+}1$ dimensions
\[
S=\frac{A}{4}\,f_Q\big|{r_2},\qquad A=\int_{0}^{2\pi}\!\!\sqrt{g_{\xi\xi}}\,d\xi=2\pi r_2.
\]
 At fixed charges/parameters, the heat capacity is
\[
C \equiv T\Big(\frac{\partial S}{\partial T}\Big)=T\,\frac{dS/dr_2}{dT/dr_2}.
\]}
\begin{figure}
\centering
\subfigure[~The behavior of the temporal component of the metric given by Eq. (\ref{metric})]{\label{fig:R1}\includegraphics[scale=0.21]{JFRMMM_KDLIAT_metra.eps}}
\subfigure[~The behavior of Eq. (\ref{metric})]{\label{fig:R}\includegraphics[scale=0.21]{JFRMMM_KDLIAT_metrb2.eps}}
\subfigure[~The behavior of temperature given by Eq. (\ref{kGR})]{\label{fig:temp1}\includegraphics[scale=0.21]{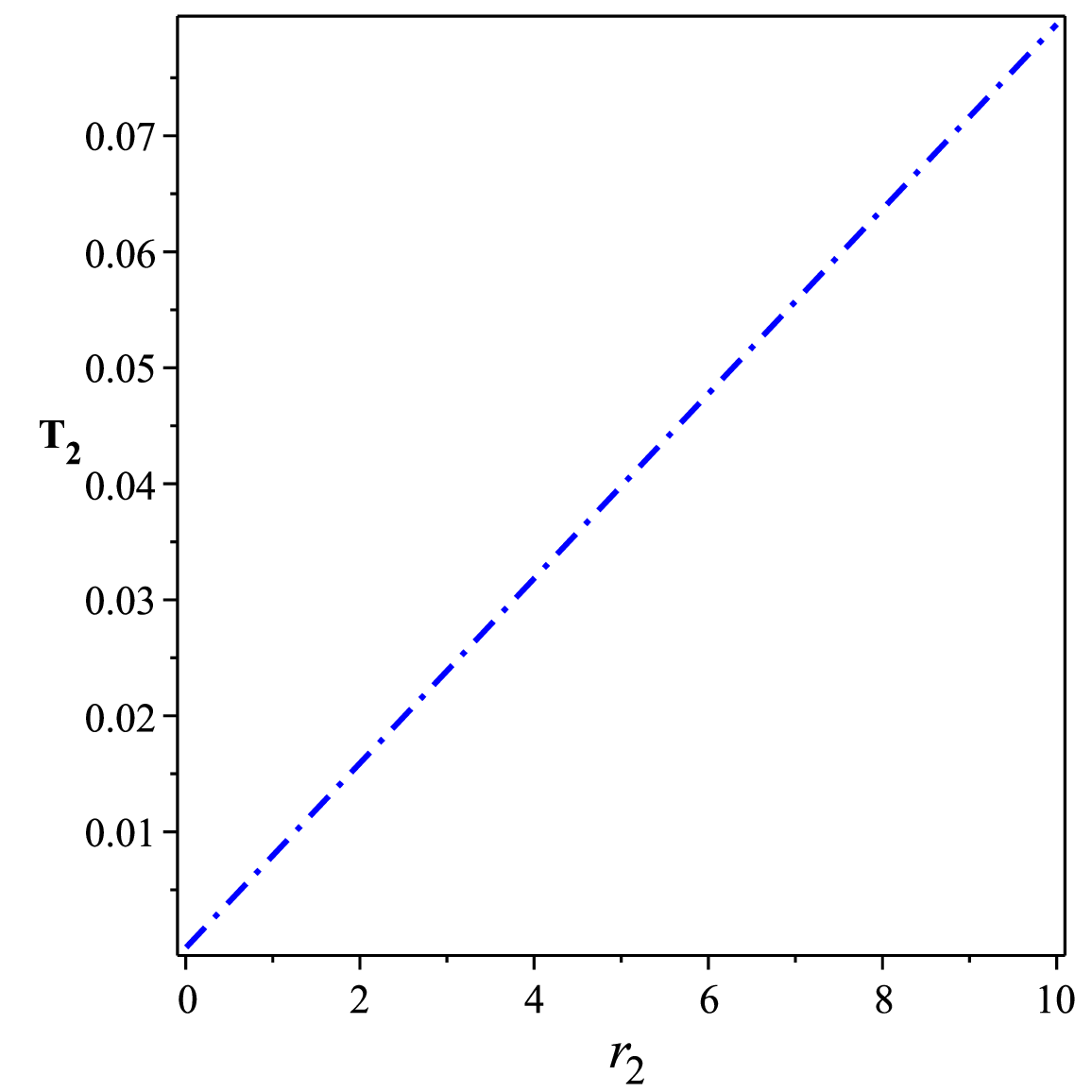}}
\subfigure[~The behavior of entropy given by Eq. (\ref{ent1})]{\label{fig:ent1}\includegraphics[scale=0.21]{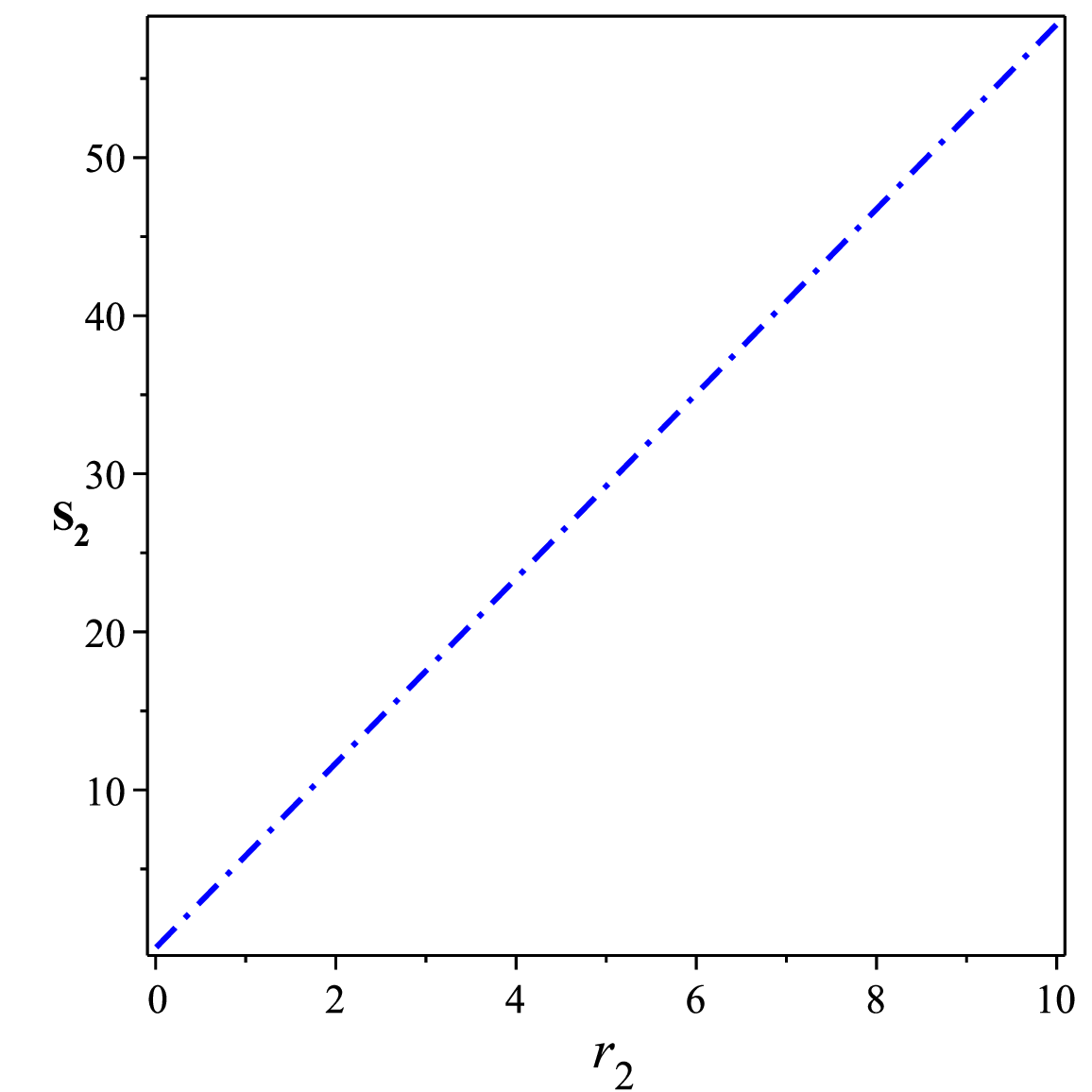}}
\subfigure[~The variation of the heat capacity defined in Eq.~(\ref{heat1})]{\label{fig:heat}\includegraphics[scale=0.21]{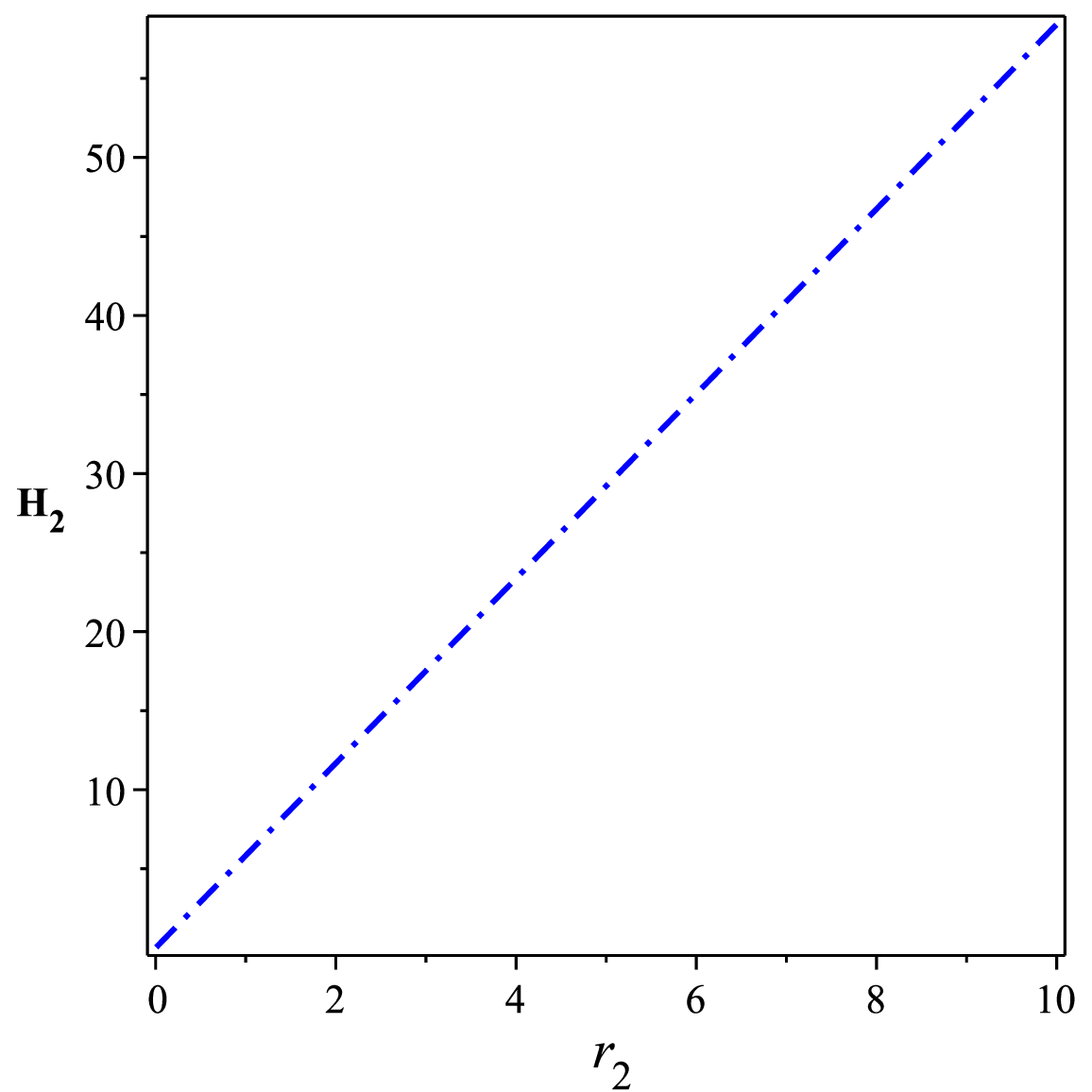}}
%
\caption[figtopcap]{\small{Panel~\subref{fig:R1} illustrates the behavior of the temporal component of the metric obtained from Eq.~(\ref{metric}), while panel~\subref{fig:R} displays the corresponding spatial component of the same metric. Panel~\subref{fig:temp1} presents the Hawking temperature calculated from Eq.~(\ref{kGR}). The entropy behavior derived from Eq.~(\ref{ent1}) is shown in panel~\subref{fig:ent1}, whereas panel~\subref{fig:heat} depicts the heat capacity obtained from Eq.~(\ref{heat1}). All curves are generated by scanning the physical outer horizon $r_2$, defined as the largest root of $k(r)=0$. In these plots, the parameters are fixed to $\Lambda=0.1$ and $m=1$, while the remaining parameters are chosen according to the set specified in Sec.~IV, ensuring that the Hawking temperature satisfies $T(r_2)>0$. In particular, the parameters are taken as $\Lambda=0.1$, $m=1$, and $c_2=10^{3}$.}}
\label{Fig:2}
\end{figure}

Fig.~\ref{Fig:1} \subref{fig:temp1} shows the behavior of the Hawking temperature described by Eq.~(\ref{kGR}), indicating that the temperature is always positive.
The semiclassical Bekenstein--Hawking entropy of the horizons is given by
\begin{align}\label{ent1}
&S(r_2)= 2\,\pi\, \left(2c_2{}^{2}- {r_2}^{2}c_3-c_4\,r_2 \right)\frac{e^{-4\Upsilon(r_2)}}{r_2\Lambda}
\,.
\end{align}
 Figure~\ref{Fig:1}\subref{fig:ent1} displays the entropy profile, demonstrating that the black hole solution corresponding to Eq.~(\ref{ent1}) yields a positive entropy. The heat capacity is then given by \cite{Zheng:2018fyn,Nashed:2019yto,Nashed:2023pxd,Kim:2012cma,Nashed:2021pah}
\begin{align} \label{heat1}
&H(r_2)=\frac{2\,\pi e^{-4\Upsilon(r_2)} \left( 2\,{c_2}^{2} -c_3{r_2}^{2}-r_2c_4\right)\left( c_3{r_2}^{2}+2r_2c_4+2\,{c_2}^{2} \right)}{r_2 \Lambda\,\left( c_3\,{r_2}^{2}+2\,{c_2}^{2} \right)}\,.
\end{align}
 We depict the behavior of the heat capacity given by Eq. (\ref{heat1})   in Fig.~\ref{Fig:1} \subref{fig:heat} which has a positive pattern ensuring that the solution under consideration is a stable one.
\section{Topological stability analysis}\label{S5}

{ The generalized free energy provides a thermodynamic potential that governs the equilibrium configurations of the system. In this formulation, the parameter $\tau$ plays the role of an effective inverse temperature associated with the Euclidean time periodicity of the black hole geometry. The equilibrium (on-shell) black hole configuration is obtained when the generalized free energy is stationary with respect to variations of the horizon radius, which occurs when $T - 1/\tau = 0$. This condition ensures consistency between the geometric temperature $T$ and the thermodynamic temperature $1/\tau$, signaling that the configuration satisfies the equations of motion. Furthermore, by analyzing the behavior near the fixed point, we find that $d\zeta_{r_2}/dr_2 > 0$ in Fig.~3(b)--(c), confirming that the scaling exponent takes the value $w = 1$, in agreement with the expected thermodynamic stability criterion.}

The study of topological structures in black hole thermodynamics relies on the concept of generalized off-shell free energy \cite{York:1986it}. In this formalism, a black hole with mass $M$ and entropy $S$ is placed inside a cavity that is kept at a constant temperature $1/\tau$. This configuration allows one to define a generalized free energy given by
\begin{align}\label{feild}
{\cal F}=M-\frac{S}{\tau},
\end{align}
which can also be derived from the gravitational path integral
\cite{Li:2022oup}. Notably, the free energy reduces to an on-shell quantity
only when $\tau = \beta = 1/T$, where T represents the Hawking
temperature.

	Having determined the generalized free energy, we adopt the framework introduced in \cite{Wei:2022dzw} to construct an appropriate parameter space and analyze the associated vector field. The physically relevant configurations appear at the locations where this vector field vanishes. These special points correspond precisely to the on-shell black hole states.

To characterize these states topologically, we compute their corresponding topological charge using Duan's $\zeta$-mapping description of topological currents \cite{Duan:1979ucg}. This charge serves as an invariant quantity that distinguishes different on-shell black hole configurations. Following the prescription of \cite{Wei:2022dzw}, the vector field $\zeta$ is introduced as
	\begin{equation}
		\zeta=(\zeta^{r_{2}},\zeta^{\theta})=(\frac{\partial {\cal F}}{\partial r_{2}},-\cot\theta \csc\theta)
	\end{equation}
	For definiteness, we restrict the angular variable to the interval $\theta\in[0,\pi]$. The vector field is therefore defined over the $(\theta,r_{2})$ plane. At the endpoints $\theta=0$ and $\theta=\pi$, the component $\zeta^{\theta}$ diverges, causing the vectors near these boundaries to orient vertically outward. The condition for a vanishing vector field corresponds to $\tau=1/T$ \cite{Wei:2021vdx}. This requirement is fulfilled only when $\theta=\pi/2$.

We now introduce the corresponding topological current as
	\begin{equation}
		j^\mu=\frac{1}{2 \pi} \epsilon^{\mu \nu \rho} \epsilon_{a b} \partial_\nu n^a \partial_\rho n^b, \quad \mu, \nu, \rho=0,1,2
	\end{equation}
	The partial derivative operator is denoted by $\partial_\nu=\frac{\partial}{\partial x^\nu}$, where the coordinate vector is chosen as $x^\nu=(\tau,r_2,\theta)$. We introduce a normalized vector field defined by $n^a=\zeta^a/\|\zeta\|$ with indices $a=1,2$. From the definition of the topological current $j^\mu$, it follows directly that the current satisfies the conservation condition $\partial_\mu j^\mu=0$. In this formulation, the parameter $\tau$ plays the role of an evolution parameter describing the development of the topological defect.

To express the current in a more useful form, we make use of the Jacobian tensor
\[
\epsilon^{ab}J^\mu(\zeta/x)=\epsilon^{\mu\nu\rho}\partial_\nu\zeta^a\,\partial_\rho\zeta^b ,
\]
together with the Green-function identity of the two-dimensional Laplacian
\[
\Delta_{\zeta^a}\ln\|\zeta\|=2\pi\,\delta^2(\zeta).
\]
Using these relations, the topological current can be written as
	\begin{equation}
		j^\mu=\delta^2(\zeta) J^\mu\left(\frac{\zeta}{x}\right) .
	\end{equation}
	Nontrivial values of the current $j^\mu$ arise exclusively at the zeros of the vector field $\zeta^a(x^i)$. Denoting the $i$-th zero by $\vec{x}=\vec{z}_i$, the density associated with the topological current is written as \cite{density}
	\begin{equation}
		j^0=\sum_{i=1}^N \beta_i \eta_i \delta^2\left(\vec{x}-\vec{z}_i\right),
	\end{equation}
Here $\beta_i$ denotes the Hopf index, which characterizes the winding multiplicity of the mapping $\zeta^a$ around the zero $z_i$ as the coordinates $x^\mu$ trace a closed path surrounding that point. As a counting number of windings, $\beta_i$ is necessarily positive. The symbol $\eta_i$ represents the Brouwer degree and is determined by the sign of the Jacobian evaluated at the zero,
$\eta_i=\mathrm{sign}\!\left[J^0(\zeta/x)\right]_{z_i}=\pm1$.
For a specified parameter domain $\Sigma$, the total topological charge can then be written as
	\begin{equation}
		W=\int_{\Sigma} j^0 d^2 x=\sum_{i=1}^N \beta_i \eta_i=\sum_{i=1}^N w_i,
		\label{W}
	\end{equation}
	Here $w_i$ represents the winding number associated with the $i$-th zero of the vector field $\zeta$ located inside the region $\Sigma$. Since it is a topological quantity, $w_i$ remains unchanged under continuous deformations of the domain used in the calculation. In general, the zeros of $\zeta$ appear as isolated points where the Jacobian satisfies $J_0(\zeta/x)\neq0$. If instead the Jacobian vanishes, $J_0(\zeta/x)=0$, the configuration indicates a bifurcation of the defect \cite{Fu:2000pb}. According to Eq.~\eqref{W}, the total topological charge in the region $\Sigma$ is obtained by summing the winding numbers of all zeros of $\zeta$, showing that the global topological number originates from these local structures.

Using the above framework, we now turn to the BTZ black hole solution. In this case, the generalized free energy can be written as
	\begin{equation}\label{free}
		{\cal F}=\Lambda \int \! \frac{\left( {r_2}^{2}c_3-2{c_2}^{2}+r_2c_4\right) ^{2}{e^{-4\Upsilon(r_2)}}}{{r_2}^{3}}dr_2-\frac{ 2\,\pi\, \left(2c_2{}^{2}- {r_2}^{2}c_3-c_4\,r_2 \right)e^{-4\Upsilon(r_2)}}{r_2\Lambda \tau},
	\end{equation}
Using Eqs.~\eqref{kGR} and \eqref{ent1}, Eq.~\eqref{feild} follows. The associated vector field is written as
\begin{equation}
		\zeta=(\phi^{r_{2}},\zeta^{\theta})=\left[\Lambda \int \! \frac{\left( {r_2}^{2}c_3-2{c_2}^{2}+r_2c_4\right) ^{2}{e^{-4\Upsilon(r_2)}}}{{r_2}^{3}}dr_2-\frac{ 2\,\pi\, \left(2c_2{}^{2}- {r_2}^{2}c_3-c_4\,r_2 \right)e^{-4\Upsilon(r_2)}}{r_2\Lambda \tau},-\cot\theta \csc\theta\right].
	\end{equation}
	
	The zeros of $\zeta$ are determined from the equation $\zeta=0$, giving the relation
	\begin{equation}\label{cons}
		\tau=\frac{2\,\pi \, \left( 2\,{c_2}^{2}-{r_2}^{2}c_3-r_2c_4 \right) {e^{-4\Upsilon(r_2)}}}{ \left( \int \! \left( {r_2}^{4}{c_3}^{2}+2\,{r_2}^{3}c_3\,c_4-4\,{r_2}^{2}c_3\,{c_2}^{2}+{r_2}^{2}{c_4}^{2}-4\, r_2c_4\,{c_2}^{2}+4\,{c_2}^{4} \right) \frac{e^{-4\Upsilon(r_2)}}{r_2{}^3}{dr_2} \right){\Lambda}^{2}{r_2}}, \quad \theta=\frac{\pi}{2}.
	\end{equation}

\begin{figure}
		\centering
		\subfigure[~Curve of zeros points of $T=\frac{1}{\tau}$]{\label{fig2:ONE}\includegraphics[scale=0.28]{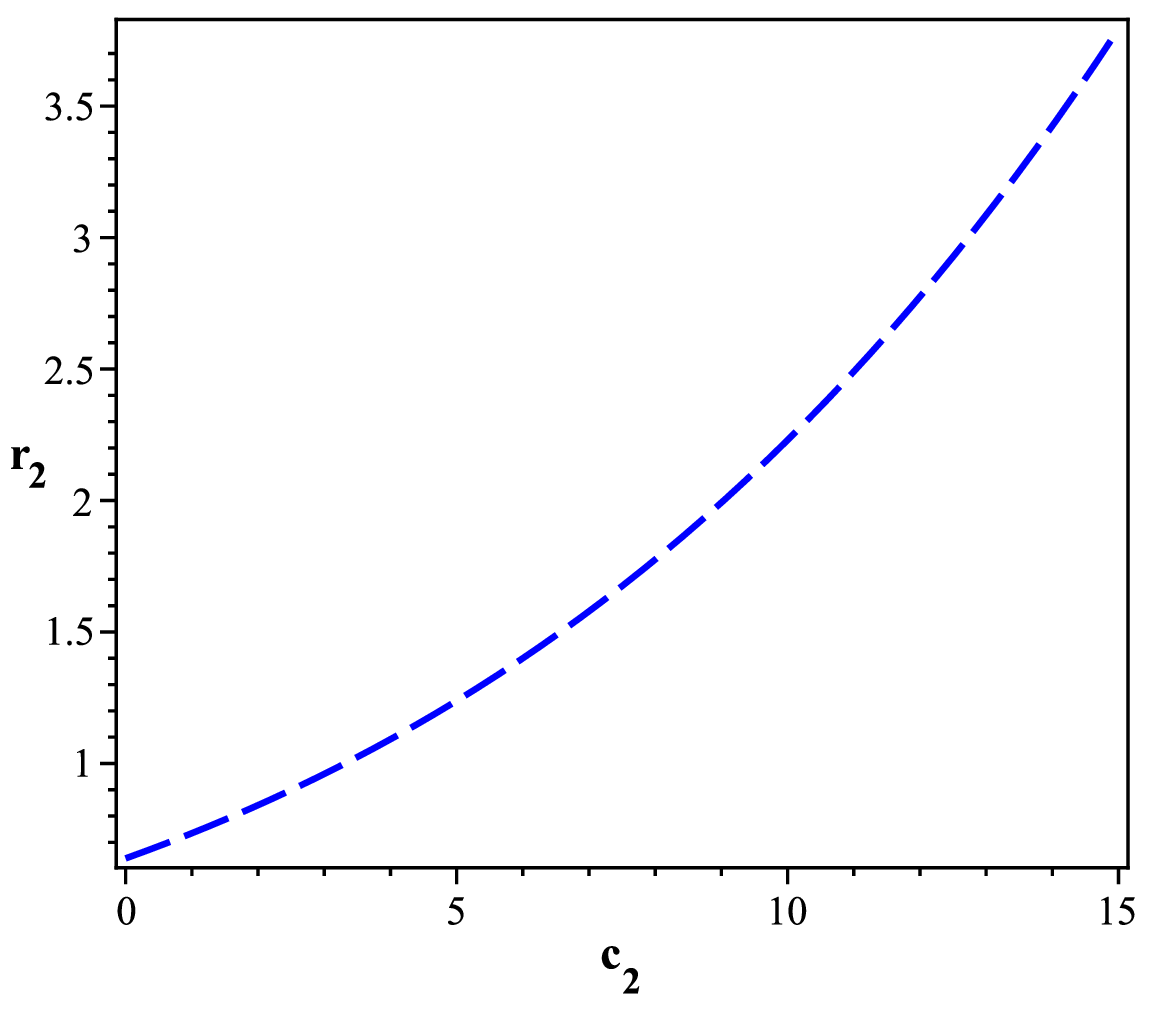}}
        \subfigure[~$\frac{d\zeta_1^{r_2}}{dr_2}$ at  solution of $T=\frac{1}{\tau}$ for $c_2>0$ ]{\label{fig2:two}\includegraphics[scale=0.25]{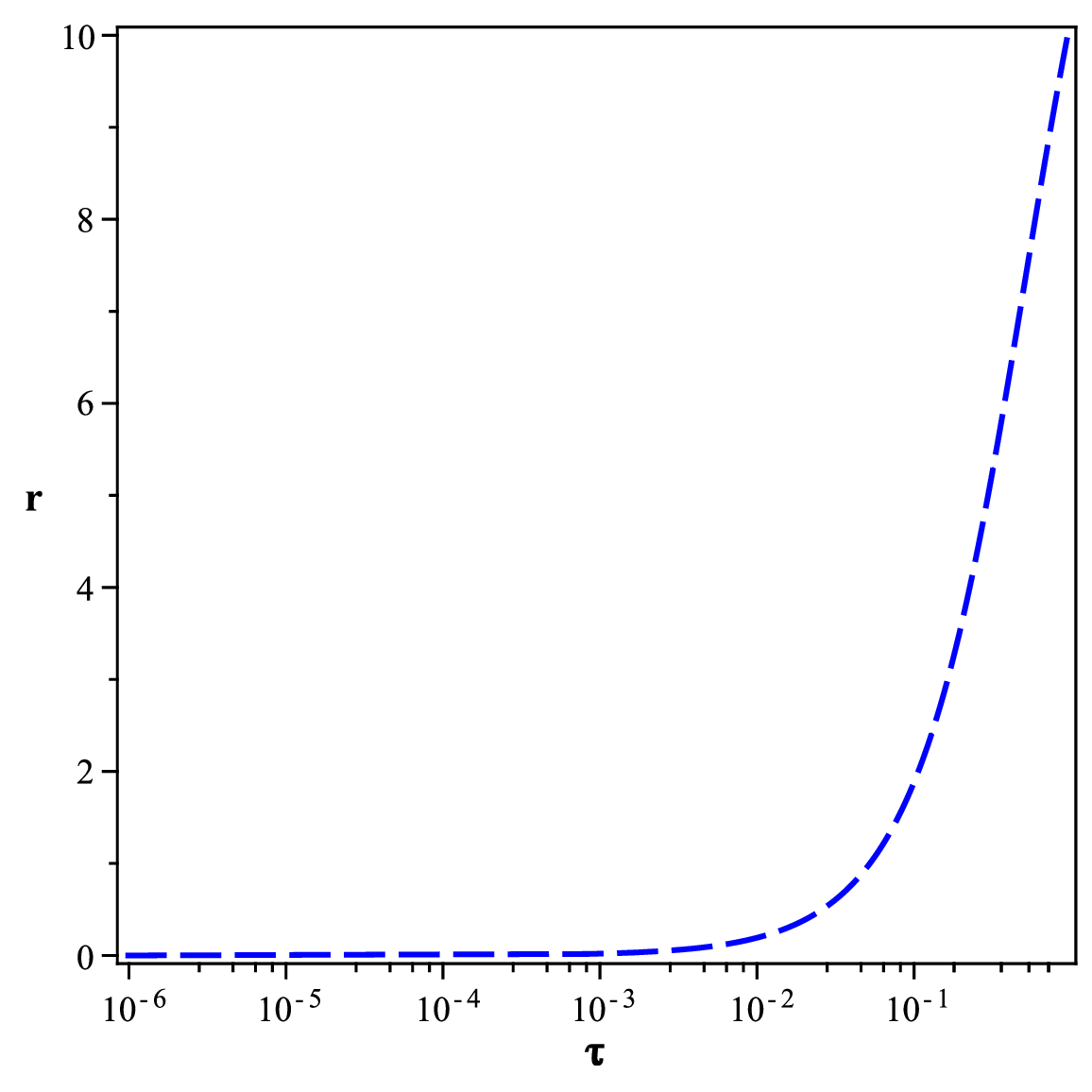}}
\subfigure[~illustrates the vector field $\frac{\zeta^r}{|\zeta|}$ with one single fixed point]{\label{fig2:three}\includegraphics[scale=0.25]{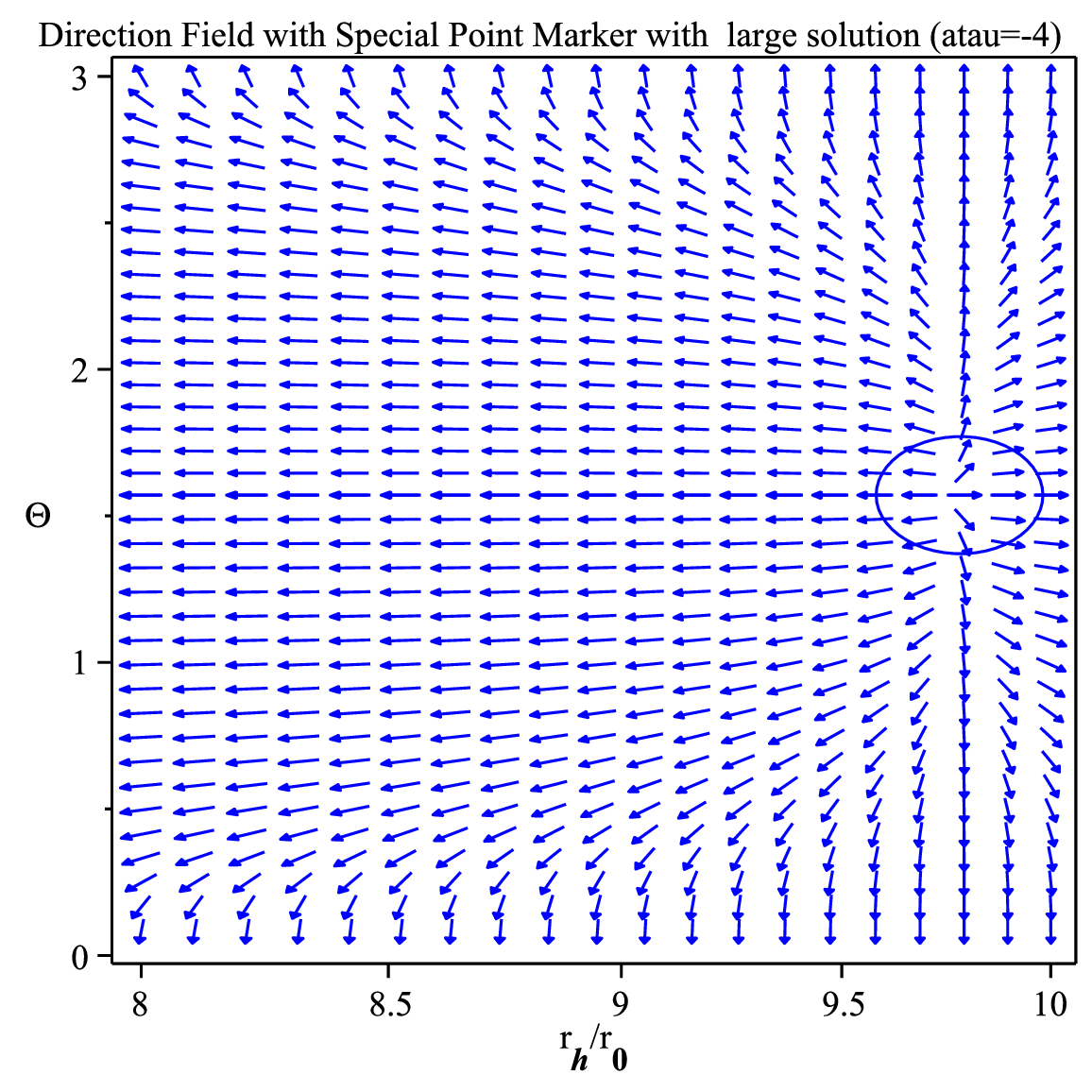}}
		\caption{Fig.~\ref{fig2}\subref{fig2:ONE} shows $\frac{d\zeta^r}{dr}$ at  solution of $T-\frac{1}{\tau}$.
Fig.~\ref{fig2}\subref{fig2:two} presents the curve corresponding to the zero points of $T-\frac{1}{\tau}=0$.
Fig.~\ref{fig2}\subref{fig2:three} illustrates the vector field $\frac{\zeta^r}{|\zeta|}$ with one single fixed point.
}
\label{fig2}
\end{figure}
In Fig.~\ref{fig2}\subref{fig2:ONE} the fixed point is located at $(9.77,\frac{\pi}{2})$ with $w=1$.
In Fig.~\ref{fig2}\subref{fig2:two}, the zero points of $T-\frac{1}{\tau}=0$ are obtained for $c_2=11$, $c_3=1$, $c_4=2$, and $\tau=0.952$.
In Fig.~\ref{fig2}\subref{fig2:ONE}, $\frac{d\zeta^r}{dr}$ also represents a solution of $T-\frac{1}{\tau}=0$ when $c_2=11$, $c_3=1$, and $c_4=14.2$.  We plot the unit vector field in Fig. \ref{fig2:three} with $c_2=11$, $c_3=c_4$.
	
	Let
\[
\zeta=\left[\frac{dS}{dr}\left(T-\frac{1}{\tau}\right),\; -\cot\theta \csc\theta\right].
\]
The black hole state that satisfies \(\zeta=0\) corresponds to the conditions
\[
T-\frac{1}{\tau}=0, \qquad \theta=\frac{\pi}{2}.
\]

To classify the winding number of this state for different choices of the constants
\((\tau, c_2, c_3, c_4)\), the parameters must first satisfy certain constraints.
Specifically, from Eq.~\eqref{deff} we require
\[
8c_2c_3 + c_4^2 > 0,
\]
since the function \(\tanh^{-1}\) has the domain \((-1,1)\), which forces the radial coordinate \(r\) to remain within this interval. To ensure that \(r \in (0,L)\) for some bounded value \(L\), we impose the conditions \(c_3>1\) and \(c_2, c_4 > 0\) and $c_4<\frac{2c_2^2-c_3r^2}{r}$.

From the condition \(T-\tfrac{1}{\tau}=0\), there exists only one positive solution which yields
 \begin{align}
 r=\frac{\sqrt{c_4^2\tau^2+8c_4\tau\pi+16\pi^2+8c_3c_2^2\tau^2}-4\pi-c_4\tau}{c_3\tau}\,,
 \end{align}
 which implies that the winding number is either \(w=1\) or \(w=-1\). Consequently, the solution depends on the bounded interval \((0,L)\). If the root lies outside this interval, additional restrictions on \(\tau\) are necessary. In particular, we find that \(\tau<0\) is required to keep the solution within \((0,L)\).

In this case, the winding number will be \(1^+\) if the sign of \(\zeta^r\) changes as \((-,+)\) around the zero, and \(1^-\) if it changes as \((+,-)\). This behavior can be analyzed through the sign of \(\tfrac{d\zeta^r}{dr}\) at the solution. By examining \(\tfrac{d\zeta^r}{dr}\) as a function of \(c_4\) for various \((\tau, c_2, c_3, c_4)\), we find that it remains positive. Hence, the winding number is always positive.

{ To summarize the study of this section: We take the components of the auxiliary field
$\zeta=(\zeta_{r_2},\zeta_{\theta})=(\partial F/\partial r_2,-\cot\theta\,\csc\theta)$
encode the gradient of the generalized free energy $F(r_2,\theta)$ on the $(r_2,\theta)$ plane.
Here $\zeta_{r_2}$ measures the radial thermodynamic response, while
$\zeta_{\theta}$ regularizes the field on the $S^2$ domain required by the $\phi$-mapping construction.
This choice ensures that the corresponding topological current
$j^{\mu}=\tfrac{1}{2\pi}\varepsilon^{\mu\nu\rho}\varepsilon^{ab}\partial_{\nu}n_a\partial_{\rho}n_b$
is divergence-free and that its zeros coincide with stationary points of $F$.

Physically, the fixed point $\zeta=0$ corresponds to
$(\partial F/\partial r_2)=0$ and $\theta=\pi/2$, i.e.\ $\tau=1/T$,
which marks an \emph{on-shell} thermodynamic equilibrium where the off-shell free energy is extremized.
In this sense, each zero of $\zeta$ represents a stable black-hole configuration in thermal equilibrium.

For the parameter set used in Fig.~3
($c_2=11$, $c_3=1$, $c_4=2$, $\tau\simeq0.95$), the local Jacobian
$J^0(\zeta/x)=\partial(\zeta_{r_2},\zeta_{\theta})/\partial(r_2,\theta)$
is positive near the zero point because $d\zeta_{r_2}/dr_2>0$ and
$\zeta_{\theta}$ changes sign from negative to positive across $\theta=\pi/2$.
Consequently, the circulation of $\zeta$ defines a single positively oriented loop,
yielding a winding number $w=+1$.
This confirms that the spacetime possesses one topologically stable thermodynamic branch.}

\subsection{Multi-horizons spacetime}\label{S661}
\begin{figure*}
\centering
\subfigure[~Multi horizons for the spacetime \eqref{sol} using different values of $m$]{\label{fig3:a}\includegraphics[scale=0.28]{JFRMMM_KDLIAT_metrbM}}
        \subfigure[~The effective potential given by Eq.~\eqref{eq:Veff}]{\label{fig3:c}\includegraphics[scale=0.28]{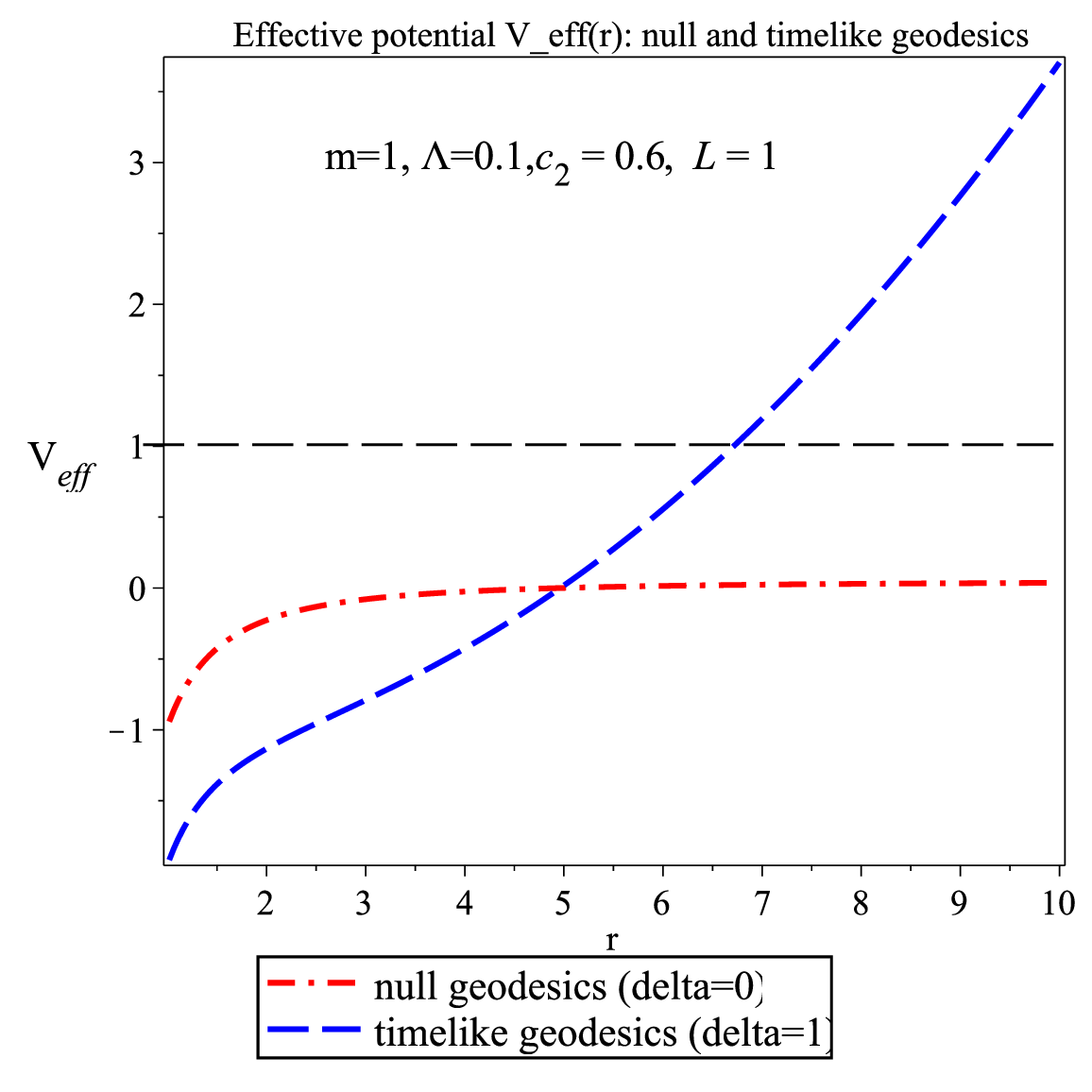}}
\caption{The three horizons of \eqref{sol}; Fig.~\ref{Fig:3}~\subref{fig3:a} for different values of $M$;
Fig.~\ref{Fig:3}~\subref{fig3:c}  the effective potential given by Eq.~\eqref{eq:Veff} when $\delta=0$ and $\delta=1$. }
\label{Fig:3}
\end{figure*}

In this section we are going discuss if the spacetime \eqref{sol} can create multi-horizon black holes or not. As we discuss in the previous section Sec. \ref{S4}  that we can reproduce two horizons when we give fixed values for the gravitational mass and the other parameters that characterize the solution. From this calculation we show that if $m=0.3$ we can create three horizons and if $m=0.1$ then two of the horizons are coincides and if $m=0.03$ then we get one  singularity. Now we are are going to  assume certain value  of $M$ and change the values $c_4$. In this case as Fig. \ref{Fig:3} shows that if $c_4=-0.3$ we can get three horizons and  if $c_4=0.254$ two of these horizons coincide and and for $c_4=0.24$ we only have one horizon. It is important to stress that if we use the numerical values of the multi horizons spacetime we can create similar  behavior for the thermodynamical quantities presented in Sec. \ref{S4}.
{ To highlight how the parameters control the causal structure, Fig.~\ref{Fig:3}\subref{fig3:a}
shows $k(r)$ for several values of the mass parameter $m$ and of the non-metricity coupling $c_4$.
Increasing $m$ shifts the outer horizon outward, while larger $c_4$ strengthens non-metricity effects,
producing transitions between single-horizon, double-horizon, and degenerate configurations.}
\section{Geodesic Completeness and Singularity Structure of the Solution}
To analyze the geodesic behavior of the metric given by Eq. \eqref{1}, employing the functions $k(r)$ and $k_1(r)$ defined in \eqref{sol}, we considered the geodesic Lagrangian, which reads \begin{align} \mathcal{L} = -k(r)\dot{t}^{2} + \frac{\dot{r}^{2}}{k_1(r)} + r^{2}\dot{\xi}^{2}, \end{align} subject to the constraint $\mathcal{L}=-1$ for timelike geodesics \cite{Wald:1999wa} and $\mathcal{L}=0$ for null geodesics \cite{Stuchlik:2001ap}. This Lagrangian is associated with conserved quantities, namely, the energy $E = k(r),\dot{t}$ and the radial momentum $L = r^{2}\dot{\xi}$ \cite{Hou:2021bxz}. These conserved quantities lead to the radial equation: \[ \dot{r}^{2} = k_1(r)\left( \frac{E^{2}}{k(r)} - \frac{L^{2}}{r^{2}} - \delta \right), \] where $\delta=1$ for timelike geodesics  and $\delta=0$ for null geodesics. Near the origin ($r=0$), the metric functions behave as $k(r) \sim -\frac{\alpha}{r^{2}}$ and $k_1(r) \sim -\beta r^{2}$, with $\alpha,\beta > 0$. This yields $\dot{r}^{2} \sim -\beta r^{2} \left( \frac{E^{2}r^{2}}{\alpha} - \frac{L^{2}}{r^{2}} - \delta \right)$, highlighting the significant effect of the energy and radial momentum near the singularity. For $L \neq 0$, this simplification demonstrates that the affine parameter needed to reach $r=0$ is finite, as previously observed \cite{Emtsova:2022uij}. For $L=0$, the equation reduces to $\dot{r}/r \sim \sqrt{\beta\delta}$, leading to the solution $r(\tau)\sim e^{\pm\tau}$, indicating that timelike geodesics terminate at $r=0$ in finite proper time \cite{Turner:2020gxo}.

{ With $E=k(r)\,\dot t$ and $L=r^{2}\dot\xi$, the radial equation
\begin{equation}
\dot r^{2}=k_{1}(r)\!\left[\frac{E^{2}}{k(r)}-\frac{L^{2}}{r^{2}}-\delta\right],\qquad
\delta=\begin{cases}1 & \text{(timelike)},\\[2pt] 0 & \text{(null)}.\end{cases}
\label{eq:rdot}
\end{equation}
It is convenient to define the effective potential
\begin{equation}
V_{\mathrm{eff}}(r;L,\delta):=k(r)\!\left(\frac{L^{2}}{r^{2}}+\delta\right),
\quad\text{so that turning points satisfy}\quad E^{2}=V_{\mathrm{eff}}(r).
\label{eq:Veff}
\end{equation}
When $k_{1}(r)>0$ (e.g.\ outside the outer horizon), the \emph{allowed} region obeys $E^{2}\ge V_{\mathrm{eff}}$; when $k_{1}(r)<0$ (inside), the inequality reverses. Using the near-center behavior derived in Sec.~VII, $k(r)\sim -\alpha/r^{2}$ and $k_{1}(r)\sim -\beta r^{2}$ with $\alpha,\beta>0$, one finds
\begin{equation}
V_{\mathrm{eff}}(r)\sim -\frac{\alpha L^{2}}{r^{4}}-\frac{\alpha\,\delta}{r^{2}}\to -\infty\quad (r\to0),
\end{equation}
and the bracket in \eqref{eq:rdot} behaves as $\frac{E^{2}}{k}-\frac{L^{2}}{r^{2}}-\delta\sim -\frac{L^{2}}{r^{2}}-\delta+O(r^{2})<0$ for $L\neq0$.
Since $k_{1}(r)<0$ near the center, $\dot r^{2}>0$ and geodesics with $L\neq0$ reach smaller $r$. For $L=0$ (radial), $\dot r^{2}\simeq \beta\,\delta\,r^{2}$, hence $r(\tau)\sim e^{\pm\tau}$ for timelike curves, showing finite proper time to $r=0$. Therefore, both null and timelike geodesics can reach the central singularity.}

{ The null case of the effective potential as shown in Fig. \ref{Fig:3}\subref{fig3:c}  displays the characteristic centrifugal barrier that prevents photons from approaching the origin,
whereas the timelike potential includes an additional constant shift that allows massive particles to penetrate deeper into the gravitational well.
For large $r$, both potentials rise monotonically, reflecting the asymptotically AdS nature of the spacetime.
These behaviors support the completeness discussion in Sec.~VII, confirming that both null and timelike trajectories can reach the central singularity in finite affine or proper time, in full agreement with the analytical results.}
\section{Discussion }
In this paper, we constructed an exact charged black hole configuration with static spherical symmetry in $(2+1)$-dimensional spacetime within the framework of $f(Q)$ gravity. The derivation was carried out without imposing extra constraints on the gravitational field equations. The obtained geometry can be viewed as a modification of the charged Banados--Teitelboim--Zanelli (BTZ) black hole solution \cite{Banados:1992wn,Banados:1992gq}, where the differences originate from additional contributions associated with the non-metricity scalar.

In the particular case where the function satisfies $f(Q)=\mathrm{const.}$, the solution naturally reduces to the standard BTZ black hole of General Relativity. For a general functional dependence on $Q$, however, the resulting spacetime exhibits more intricate geometric and physical properties. This demonstrates that $f(Q)$ gravity provides a consistent framework for describing lower-dimensional black hole solutions and offers extensions beyond the standard predictions of GR \cite{DAmbrosio:2021zpm}.

Curvature and non-metricity invariants confirm that the central singularity in the $f(Q)$ solution is stronger than in GR, with the Ricci scalar diverging as $r\to 0$, unlike the constant value in GR. This divergent behavior, influenced by the inclusion of non-metricity corrections, suggests a regularization effect related to the modified geometry. Furthermore, the metric asymptotically approaches anti-de Sitter (AdS) space, guaranteeing compatibility with the AdS/CFT correspondence.

The analysis of  non-metricity and curvature scalars reveals that the central singularity in the $f(Q)$ solution is considerably strong compared to GR. Specifically, while the Kretschmann scalar in GR-type solutions diverges strongly near $r=0$, the non-metricity corrections lead to milder divergences, suggesting a possible regularization effect of modified geometry. Furthermore, the metric asymptotically behaves as anti-de Sitter (AdS), ensuring compatibility with the AdS/CFT correspondence \cite{Maldacena:1997re,Witten:1998qj}, which makes the solution relevant for holographic studies of lower-dimensional black holes.

 { From a thermodynamical perspective, the black hole exhibits standard Hawking radiation with a positive temperature and entropy, while its heat capacity remains strictly positive, ensuring local thermodynamical stability. This behavior is distinct from many GR-based black holes, where instabilities typically arise in certain parameter regimes \cite{Zheng:2018fyn}. The stability of the solution further strengthens the physical viability of the derived geometry.}

 The topological characterization, based on Duan's $\phi$-mapping current theory \cite{Duan:1979ucg}, demonstrates that the black hole solution belongs to a single topological class with winding number $w=1$. This indicates that the thermodynamical phase space of the black hole in $f(Q)$ gravity is topologically stable, in agreement with recent studies of black hole thermodynamic topology \cite{Wei:2022dzw}. Moreover, we have shown that depending on the choice of parameters, the solution can admit multiple horizons, with configurations ranging from three horizons to extremal or naked singular states. This enriches the phenomenological landscape of $(2+1)$-dimensional gravity and suggests new avenues for investigating horizon thermodynamics in modified gravity.

 In conclusion, our findings show that $f(Q)$ gravity provides a natural generalization of the charged BTZ solution, strongest central singularities compared with the squared of Ricci tensor and Ricci tensor of the GR black holes, ensures thermodynamical stability, and admits a consistent topological classification. These features underline the potential of $f(Q)$ gravity to serve as a promising framework for exploring quantum aspects of gravity and the holographic principle in lower dimensions. Future directions include extending the analysis to rotating configurations, higher-dimensional analogues, and coupling with matter fields to further probe the astrophysical implications of $f(Q)$ gravity.
\newpage
{ \appendix
\section*{Appendix~A: The details forms of the field equations}

Using Eq.~\eqref{dfg} in Eq.~\eqref{fe1} we get:
\begin{eqnarray}\label{fe3}
&& \zeta_t{}^{t}\equiv0=\frac {1}{2 k\left[ k''k_1 rk + k'  \left\{k  \left( k'_1 r -k_1  \right) -  k'  k_1 r  \right\}  \right] ^{2}}\left[2k^{3}{r}^{2}k_1^{2}f' k''' +f  k''^{2}k_1^ {2}{r}^{2}k^{2}+2r \left[ -k^{2}f'' rk_1 -r k'^{2}f k_1\right.\right.\nonumber\\
 &&\left.\left.+k  \left\{f  \left(  k'_1 r-k_1  \right)   -\frac{7}2k_1 f' r\right\} k' +\frac{1}2 f'  \left(3 k'_1 r -4k_1 \right)  k^{2} \right] k_1 k k'' + \left\{ 2rk_1  \left( k' k_1 r+k  \left\{k_1 -k'_1 r \right\}  \right)  k^{2} f''\right.\right.\nonumber\\
 &&\left.\left. +2 k^{3}k_1 f' k''_1{r}^{2}+f k_1^{2}{r}^{2}k'^{3}+ \left( 5k_1f' r-2f  \left\{ k'_1 r -k_1  \right\}  \right) rk_1 k  k'^{2}+ \left[  \left( 5r k_1^{2}-4  k'_1  k_1 {r}^{2} \right) f' +f  \left( k'_1 r -k_1 \right) ^{2} \right] k^{2}k'\right.\right.\nonumber\\
 &&\left.\left. - k^{3} f' \left( 4k_1 + k'_1 r \right) \left(  k'_1 r -k_1 \right)  \right\} k'\right],\label{2} \nonumber\\
&&\zeta_r{}^{r}\equiv0=\frac {rk  k_1   \left( f  k  -2 \varphi'^{2}k_1   \right)k''  + \left( -rk_1  \left( f  k  -2 \varphi'^{2}k_1   \right) k'  +k   \left[ \left( 2  k_1^{2}-2k_1   k'_1 r \right)  \varphi'^{2}+k   \left\{ f  k'_1 r-k_1 \left( f  + 2f'  r \right)  \right\}  \right]  \right) k'  }{2 \left[  k''   k_1  rk  + \left\{ k   \left(   k'_1 r -k_1\right)   -k_1  rk'\right\} k'   \right] k  }\,,\nonumber\\
&&\zeta_\xi{}^{\xi}\equiv0=\frac {1}{ 4\left[  k'' k_1  rk  + \left\{ k   \left( k'_1r -k_1 \right) -k_1  rk'  \right\} k'   \right] ^{2}k  }\left[2 k_1 ^{2} k^{3} f' {r}^{3} k''' k'  -2{r}^{2} k_1 ^{2} k^{2} \left( -f k  +k   f' r-2 \varphi'^{2}k_1   \right) k''^{2}\right.\nonumber\\
 &&\left.-k_1 k  r \left\{2  k^{2} k_1  {r}^{2}f''  +k_1  r \left(8 \varphi'^{2}k_1  +3k   f' r+4f k   \right) k'  -\left[ r \left( 4f k  +k   f' r+8 \varphi'^{2}k_1   \right) k'_1  -4k_1   \left( 2 \varphi'^{2}k_1  +f k  +k   f' r \right)  \right]k   \right\}  k' k'' \right.\nonumber\\
 &&\left. + k'^2 \left\{ 2k_1   \left[ k_1  rk'  +k   \left( k_1  - k'_1 r \right)  \right] k^2r^2f''  +2k_1   k^3 f' r^3k''_1  + k_1 ^2r^2 \left( 4 \varphi'^2k_1  +3k   f' r+2f k   \right) k'^2- \left\{ r \left(k   f' r+2f k +4 \varphi'^2 k_1   \right) k'_1\right.\right.\right.\nonumber\\
 &&\left.\left.\left.  -2k  f  k_1  -\frac{5}2k   f' r k_1 -4 \varphi'^{2} k_1 ^{2} \right\}2k_1k  rk'  -\frac{k^{2}}2  \left\{ r \left(k   f' r -4 \varphi'^{2}k_1 -2f k   \right) k'_1  +2k  f  k_1 +4 \varphi'^{2} k_1 ^{2}+6k   f' rk_1   \right\}  \left(  k'_1 r-k_1 \right)  \right\}\right]
\,, \nonumber\\
&& \mbox {and finally the equation of the electric field takes the form} \,\,\, 0= \frac {2\,rk_1  \varphi''  k +r \varphi'  k'_1 k -r \varphi'  k_1 k' +2\,\varphi' k k_1 }{2k^{2}r}.
\end{eqnarray}}
{
\appendix
\section*{Appendix~B: Detailed Derivation of the Field Equations in $(2+1)$-Dimensional $f(Q)$ Gravity}

In this appendix, we present the full derivation of the field equations and the functional dependence of $f$, $f_Q$, and $f_{QQ}$ that lead from the action~(\ref{action1}) to Eqs.~(\ref{fe1})--(\ref{sol}) in the main text.

Using Eq.~(\ref{1}) and the non-metricity tensor, i.e., $Q_{\gamma\mu\nu}=-\partial_{\gamma} g_{\mu\nu}$
we compute its non-vanishing components that take the following form,
\[
Q_{rtt} = k'(r), \qquad Q_{rrr} = -\frac{k_{1}'(r)}{k_{1}^{2}(r)}, \qquad Q_{r\xi\xi} = -2r, \qquad \mbox{where a prime denotes $d/dr$.}
\]
Using the disformation tensor, $
L^{\gamma}{}_{\mu\nu}$, given by Eq. \eqref{dis}, and inserting the above components, we obtain the non-metricity scalar  by Eq.~(\ref{Qc}) in the main text. For the Maxwell field, we adopt the potential one-form $\phi = \phi(r)\,dt$, yielding the field tensor $F_{tr}=-F_{rt}=\phi'(r)$ where corresponding energy-momentum tensor given by \eqref{Max}. Using Eq.~(\ref{Qc}) in the (\ref{1st EOM}), we compute the independent components of given by (\ref{fe1}).
Since $Q$ depends only on $r$ through Eq.~(\ref{Qc}), we can write $f(Q)\equiv f(r)$, and therefore
\[
\frac{df}{dQ} = \frac{df/dr}{dQ/dr} = f_{Q}, \qquad
\frac{d^{2}f}{dQ^{2}} = \frac{d}{dQ}\!\left(\frac{df}{dQ}\right)
= \frac{d}{dr}\!\left(\frac{df/dr}{dQ/dr}\right)\!\Big/\!\left(\frac{dQ}{dr}\right).
\]
Differentiating Eq.~(\ref{Qc})  gives
\begin{equation}
\frac{dQ}{dr} = -\frac{k_{1}'k'}{r k}
-\frac{k_{1}k''}{r k}
+\frac{k_{1} (k')^{2}}{r k^{2}}
+\frac{k_{1}k'}{r^{2} k}.
\label{A6}
\end{equation}
Substituting (\ref{A6}) into the above expressions yields:
\begin{align}
f_{Q}
= -\,\frac{f'(r)}{\,dQ/dr\,}
= -\,\frac{f'(r)\,r^{2}k^{2}}
{k_{1} \big(k'' k r - (k')^{2}r + k'k\big)},\qquad
f_{QQ}
= -\,\frac{1}{(dQ/dr)^{3}}
\left[
f''(r)\left(\frac{dQ}{dr}\right)
- f'(r)\,\frac{d^{2}Q}{dr^{2}}
\right].
\label{A7}
\end{align}
Equation~(\ref{A7}) reproduces the general functional relations given in Eq.~(\ref{dfg}).

Substituting Eq.~(\ref{A7}) and the expression for $Q(r)$ into (\ref{fe1}) provides a closed system of four nonlinear ordinary differential equations in the four unknown functions $k(r)$, $k_{1}(r)$, $f(r)$, and $\phi(r)$:
\begin{equation}
\big\{\,\zeta^{t}{}_{t}=0,\;
\zeta^{r}{}_{r}=0,\;
\zeta^{\xi}{}_{\xi}=0,\;
\nabla_{\mu}F^{\mu\nu}=0\,\big\}.
\label{A9}
\end{equation}
Solving the coupled system~(\ref{A9}) yields the exact analytic expressions summarized in Eq.~(\ref{sol}).}

\subsection*{Acknowledgments}
This work was supported and funded by the Deanship of Scientific Research at Imam Mohammad Ibn Saud Islamic University (IMSIU) (grant number
IMSIU-DDRSP2502).

\begin{thebibliography}{71}
\expandafter\ifx\csname natexlab\endcsname\relax\def\natexlab#1{#1}\fi
\expandafter\ifx\csname bibnamefont\endcsname\relax
  \def\bibnamefont#1{#1}\fi
\expandafter\ifx\csname bibfnamefont\endcsname\relax
  \def\bibfnamefont#1{#1}\fi
\expandafter\ifx\csname citenamefont\endcsname\relax
  \def\citenamefont#1{#1}\fi
\expandafter\ifx\csname url\endcsname\relax
  \def\url#1{\texttt{#1}}\fi
\expandafter\ifx\csname urlprefix\endcsname\relax\def\urlprefix{URL }\fi
\providecommand{\bibinfo}[2]{#2}
\providecommand{\eprint}[2][]{\url{#2}}

\bibitem[{\citenamefont{Beltr{\'a}n~Jim{\'e}nez
  et~al.}(2019)\citenamefont{Beltr{\'a}n~Jim{\'e}nez, Heisenberg, and
  Koivisto}}]{BeltranJimenez:2019esp}
\bibinfo{author}{\bibfnamefont{J.}~\bibnamefont{Beltr{\'a}n~Jim{\'e}nez}},
  \bibinfo{author}{\bibfnamefont{L.}~\bibnamefont{Heisenberg}},
  \bibnamefont{and} \bibinfo{author}{\bibfnamefont{T.~S.}
  \bibnamefont{Koivisto}}, \bibinfo{journal}{Universe}
  \textbf{\bibinfo{volume}{5}}, \bibinfo{pages}{173} (\bibinfo{year}{2019}),
  \eprint{1903.06830}.

\bibitem[{\citenamefont{Harada}(2020)}]{Harada:2020ikm}
\bibinfo{author}{\bibfnamefont{J.}~\bibnamefont{Harada}},
  \bibinfo{journal}{Phys. Rev. D} \textbf{\bibinfo{volume}{101}},
  \bibinfo{pages}{024053} (\bibinfo{year}{2020}), \eprint{2001.06990}.

\bibitem[{\citenamefont{Aldrovandi and Pereira}(2013)}]{Aldrovandi:2013wha}
\bibinfo{author}{\bibfnamefont{R.}~\bibnamefont{Aldrovandi}} \bibnamefont{and}
  \bibinfo{author}{\bibfnamefont{J.~G.} \bibnamefont{Pereira}},
  \emph{\bibinfo{title}{{Teleparallel Gravity}: {An Introduction}}}
  (\bibinfo{publisher}{Springer}, \bibinfo{year}{2013}), ISBN
  \bibinfo{isbn}{978-94-007-5142-2, 978-94-007-5143-9}.

\bibitem[{\citenamefont{Maluf}(2013)}]{Maluf:2013gaa}
\bibinfo{author}{\bibfnamefont{J.~W.} \bibnamefont{Maluf}},
  \bibinfo{journal}{Annalen Phys.} \textbf{\bibinfo{volume}{525}},
  \bibinfo{pages}{339} (\bibinfo{year}{2013}), \eprint{1303.3897}.

\bibitem[{\citenamefont{Nester and Yo}(1999)}]{Nester:1998mp}
\bibinfo{author}{\bibfnamefont{J.~M.} \bibnamefont{Nester}} \bibnamefont{and}
  \bibinfo{author}{\bibfnamefont{H.-J.} \bibnamefont{Yo}},
  \bibinfo{journal}{Chin. J. Phys.} \textbf{\bibinfo{volume}{37}},
  \bibinfo{pages}{113} (\bibinfo{year}{1999}), \eprint{gr-qc/9809049}.

\bibitem[{\citenamefont{Adak et~al.}(2006)\citenamefont{Adak, Kalay, and
  Sert}}]{Adak:2005cd}
\bibinfo{author}{\bibfnamefont{M.}~\bibnamefont{Adak}},
  \bibinfo{author}{\bibfnamefont{M.}~\bibnamefont{Kalay}}, \bibnamefont{and}
  \bibinfo{author}{\bibfnamefont{O.}~\bibnamefont{Sert}},
  \bibinfo{journal}{Int. J. Mod. Phys. D} \textbf{\bibinfo{volume}{15}},
  \bibinfo{pages}{619} (\bibinfo{year}{2006}), \eprint{gr-qc/0505025}.

\bibitem[{\citenamefont{Adak et~al.}(2013)\citenamefont{Adak, Ali, and
  Majumdar}}]{Adak:2013vwa}
\bibinfo{author}{\bibfnamefont{D.}~\bibnamefont{Adak}},
  \bibinfo{author}{\bibfnamefont{A.}~\bibnamefont{Ali}}, \bibnamefont{and}
  \bibinfo{author}{\bibfnamefont{D.}~\bibnamefont{Majumdar}},
  \bibinfo{journal}{Phys. Rev. D} \textbf{\bibinfo{volume}{88}},
  \bibinfo{pages}{024007} (\bibinfo{year}{2013}), \eprint{1305.2330}.

\bibitem[{\citenamefont{Mol}(2017)}]{Mol:2014ooa}
\bibinfo{author}{\bibfnamefont{I.}~\bibnamefont{Mol}}, \bibinfo{journal}{Adv.
  Appl. Clifford Algebras} \textbf{\bibinfo{volume}{27}}, \bibinfo{pages}{2607}
  (\bibinfo{year}{2017}), \eprint{1406.0737}.

\bibitem[{\citenamefont{J{\"a}rv et~al.}(2018)\citenamefont{J{\"a}rv,
  R{\"u}nkla, Saal, and Vilson}}]{Jarv:2018bgs}
\bibinfo{author}{\bibfnamefont{L.}~\bibnamefont{J{\"a}rv}},
  \bibinfo{author}{\bibfnamefont{M.}~\bibnamefont{R{\"u}nkla}},
  \bibinfo{author}{\bibfnamefont{M.}~\bibnamefont{Saal}}, \bibnamefont{and}
  \bibinfo{author}{\bibfnamefont{O.}~\bibnamefont{Vilson}},
  \bibinfo{journal}{Phys. Rev. D} \textbf{\bibinfo{volume}{97}},
  \bibinfo{pages}{124025} (\bibinfo{year}{2018}), \eprint{1802.00492}.

\bibitem[{\citenamefont{De~Felice and Tsujikawa}(2010)}]{DeFelice:2010aj}
\bibinfo{author}{\bibfnamefont{A.}~\bibnamefont{De~Felice}} \bibnamefont{and}
  \bibinfo{author}{\bibfnamefont{S.}~\bibnamefont{Tsujikawa}},
  \bibinfo{journal}{Living Rev. Rel.} \textbf{\bibinfo{volume}{13}},
  \bibinfo{pages}{3} (\bibinfo{year}{2010}), \eprint{1002.4928}.

\bibitem[{\citenamefont{Sotiriou and Faraoni}(2010)}]{Sotiriou:2008rp}
\bibinfo{author}{\bibfnamefont{T.~P.} \bibnamefont{Sotiriou}} \bibnamefont{and}
  \bibinfo{author}{\bibfnamefont{V.}~\bibnamefont{Faraoni}},
  \bibinfo{journal}{Rev. Mod. Phys.} \textbf{\bibinfo{volume}{82}},
  \bibinfo{pages}{451} (\bibinfo{year}{2010}), \eprint{0805.1726}.

\bibitem[{\citenamefont{Capozziello and
  De~Laurentis}(2011)}]{Capozziello:2011et}
\bibinfo{author}{\bibfnamefont{S.}~\bibnamefont{Capozziello}} \bibnamefont{and}
  \bibinfo{author}{\bibfnamefont{M.}~\bibnamefont{De~Laurentis}},
  \bibinfo{journal}{Phys. Rept.} \textbf{\bibinfo{volume}{509}},
  \bibinfo{pages}{167} (\bibinfo{year}{2011}), \eprint{1108.6266}.

\bibitem[{\citenamefont{Nojiri and Odintsov}(2011)}]{Nojiri:2010wj}
\bibinfo{author}{\bibfnamefont{S.}~\bibnamefont{Nojiri}} \bibnamefont{and}
  \bibinfo{author}{\bibfnamefont{S.~D.} \bibnamefont{Odintsov}},
  \bibinfo{journal}{Phys. Rept.} \textbf{\bibinfo{volume}{505}},
  \bibinfo{pages}{59} (\bibinfo{year}{2011}), \eprint{1011.0544}.

\bibitem[{\citenamefont{Bengochea and Ferraro}(2009)}]{Bengochea:2008gz}
\bibinfo{author}{\bibfnamefont{G.~R.} \bibnamefont{Bengochea}}
  \bibnamefont{and} \bibinfo{author}{\bibfnamefont{R.}~\bibnamefont{Ferraro}},
  \bibinfo{journal}{Phys. Rev. D} \textbf{\bibinfo{volume}{79}},
  \bibinfo{pages}{124019} (\bibinfo{year}{2009}), \eprint{0812.1205}.

\bibitem[{\citenamefont{Yousaf et~al.}(2025{\natexlab{a}})\citenamefont{Yousaf,
  Khokhar, Alshammari, Albalahi, Abdulrahman, and Alshammari}}]{yousaf2025dark}
\bibinfo{author}{\bibfnamefont{Z.}~\bibnamefont{Yousaf}},
  \bibinfo{author}{\bibfnamefont{U.~A.} \bibnamefont{Khokhar}},
  \bibinfo{author}{\bibfnamefont{M.}~\bibnamefont{Alshammari}},
  \bibinfo{author}{\bibfnamefont{A.~M.} \bibnamefont{Albalahi}},
  \bibinfo{author}{\bibfnamefont{A.~T.} \bibnamefont{Abdulrahman}},
  \bibnamefont{and} \bibinfo{author}{\bibfnamefont{T.~S.}
  \bibnamefont{Alshammari}}, \bibinfo{journal}{Int. J. Geom. Methods Mod.
  Phys.} p. \bibinfo{pages}{10.1142/S0219887826500143}
  (\bibinfo{year}{2025}{\natexlab{a}}).

\bibitem[{\citenamefont{Nashed et~al.}(2020)\citenamefont{Nashed, El~Hanafy,
  Odintsov, and Oikonomou}}]{Nashed:2019yto}
\bibinfo{author}{\bibfnamefont{G.~G.~L.} \bibnamefont{Nashed}},
  \bibinfo{author}{\bibfnamefont{W.}~\bibnamefont{El~Hanafy}},
  \bibinfo{author}{\bibfnamefont{S.~D.} \bibnamefont{Odintsov}},
  \bibnamefont{and} \bibinfo{author}{\bibfnamefont{V.~K.}
  \bibnamefont{Oikonomou}}, \bibinfo{journal}{Int. J. Mod. Phys. D}
  \textbf{\bibinfo{volume}{29}}, \bibinfo{pages}{2050090}
  (\bibinfo{year}{2020}), \eprint{1912.03897}.

\bibitem[{\citenamefont{Alshammari
  et~al.}(2025{\natexlab{a}})\citenamefont{Alshammari, Aman
  et~al.}}]{alshammari2025mass}
\bibinfo{author}{\bibfnamefont{M.}~\bibnamefont{Alshammari}},
  \bibinfo{author}{\bibfnamefont{H.}~\bibnamefont{Aman}}, \bibnamefont{et~al.},
  \bibinfo{journal}{Int. J. Geom. Methods Mod. Phys.} p.
  \bibinfo{pages}{10.1142/S0219887826500428}
  (\bibinfo{year}{2025}{\natexlab{a}}).

\bibitem[{\citenamefont{Linder}(2010)}]{Linder:2010py}
\bibinfo{author}{\bibfnamefont{E.~V.} \bibnamefont{Linder}},
  \bibinfo{journal}{Phys. Rev. D} \textbf{\bibinfo{volume}{81}},
  \bibinfo{pages}{127301} (\bibinfo{year}{2010}), \bibinfo{note}{[Erratum:
  Phys.Rev.D 82, 109902 (2010)]}, \eprint{1005.3039}.

\bibitem[{\citenamefont{Cai et~al.}(2016)\citenamefont{Cai, Capozziello,
  De~Laurentis, and Saridakis}}]{Cai:2015emx}
\bibinfo{author}{\bibfnamefont{Y.-F.} \bibnamefont{Cai}},
  \bibinfo{author}{\bibfnamefont{S.}~\bibnamefont{Capozziello}},
  \bibinfo{author}{\bibfnamefont{M.}~\bibnamefont{De~Laurentis}},
  \bibnamefont{and} \bibinfo{author}{\bibfnamefont{E.~N.}
  \bibnamefont{Saridakis}}, \bibinfo{journal}{Rept. Prog. Phys.}
  \textbf{\bibinfo{volume}{79}}, \bibinfo{pages}{106901}
  (\bibinfo{year}{2016}), \eprint{1511.07586}.

\bibitem[{\citenamefont{Nojiri et~al.}(2017)\citenamefont{Nojiri, Odintsov, and
  Oikonomou}}]{Nojiri:2017ncd}
\bibinfo{author}{\bibfnamefont{S.}~\bibnamefont{Nojiri}},
  \bibinfo{author}{\bibfnamefont{S.~D.} \bibnamefont{Odintsov}},
  \bibnamefont{and} \bibinfo{author}{\bibfnamefont{V.~K.}
  \bibnamefont{Oikonomou}}, \bibinfo{journal}{Phys. Rept.}
  \textbf{\bibinfo{volume}{692}}, \bibinfo{pages}{1} (\bibinfo{year}{2017}),
  \eprint{1705.11098}.

\bibitem[{\citenamefont{Lu et~al.}(2019)\citenamefont{Lu, Zhao, and
  Chee}}]{Lu:2019hra}
\bibinfo{author}{\bibfnamefont{J.}~\bibnamefont{Lu}},
  \bibinfo{author}{\bibfnamefont{X.}~\bibnamefont{Zhao}}, \bibnamefont{and}
  \bibinfo{author}{\bibfnamefont{G.}~\bibnamefont{Chee}},
  \bibinfo{journal}{Eur. Phys. J. C} \textbf{\bibinfo{volume}{79}},
  \bibinfo{pages}{530} (\bibinfo{year}{2019}), \eprint{1906.08920}.

\bibitem[{\citenamefont{Beltr{\'a}n~Jim{\'e}nez
  et~al.}(2018{\natexlab{a}})\citenamefont{Beltr{\'a}n~Jim{\'e}nez, Heisenberg,
  and Koivisto}}]{BeltranJimenez:2018vdo}
\bibinfo{author}{\bibfnamefont{J.}~\bibnamefont{Beltr{\'a}n~Jim{\'e}nez}},
  \bibinfo{author}{\bibfnamefont{L.}~\bibnamefont{Heisenberg}},
  \bibnamefont{and} \bibinfo{author}{\bibfnamefont{T.~S.}
  \bibnamefont{Koivisto}}, \bibinfo{journal}{JCAP}
  \textbf{\bibinfo{volume}{08}}, \bibinfo{pages}{039}
  (\bibinfo{year}{2018}{\natexlab{a}}), \eprint{1803.10185}.

\bibitem[{\citenamefont{Yousaf et~al.}(2025{\natexlab{b}})\citenamefont{Yousaf,
  Asad et~al.}}]{yousaf2025electromagnetic}
\bibinfo{author}{\bibfnamefont{Z.}~\bibnamefont{Yousaf}},
  \bibinfo{author}{\bibfnamefont{H.}~\bibnamefont{Asad}}, \bibnamefont{et~al.},
  \bibinfo{journal}{Int. J. Geom. Methods Mod. Phys.} p.
  \bibinfo{pages}{2550145} (\bibinfo{year}{2025}{\natexlab{b}}).

\bibitem[{\citenamefont{Lazkoz et~al.}(2019)\citenamefont{Lazkoz, Lobo,
  Ortiz-Ba{\~n}os, and Salzano}}]{Lazkoz:2019sjl}
\bibinfo{author}{\bibfnamefont{R.}~\bibnamefont{Lazkoz}},
  \bibinfo{author}{\bibfnamefont{F.~S.~N.} \bibnamefont{Lobo}},
  \bibinfo{author}{\bibfnamefont{M.}~\bibnamefont{Ortiz-Ba{\~n}os}},
  \bibnamefont{and} \bibinfo{author}{\bibfnamefont{V.}~\bibnamefont{Salzano}},
  \bibinfo{journal}{Phys. Rev. D} \textbf{\bibinfo{volume}{100}},
  \bibinfo{pages}{104027} (\bibinfo{year}{2019}), \eprint{1907.13219}.

\bibitem[{\citenamefont{Mandal et~al.}(2020{\natexlab{a}})\citenamefont{Mandal,
  Wang, and Sahoo}}]{Mandal:2020buf}
\bibinfo{author}{\bibfnamefont{S.}~\bibnamefont{Mandal}},
  \bibinfo{author}{\bibfnamefont{D.}~\bibnamefont{Wang}}, \bibnamefont{and}
  \bibinfo{author}{\bibfnamefont{P.~K.} \bibnamefont{Sahoo}},
  \bibinfo{journal}{Phys. Rev. D} \textbf{\bibinfo{volume}{102}},
  \bibinfo{pages}{124029} (\bibinfo{year}{2020}{\natexlab{a}}),
  \eprint{2011.00420}.

\bibitem[{\citenamefont{Beltr\'an~Jim\'enez
  et~al.}(2020)\citenamefont{Beltr\'an~Jim\'enez, Heisenberg, Koivisto, and
  Pekar}}]{BeltranJimenez:2019tme}
\bibinfo{author}{\bibfnamefont{J.}~\bibnamefont{Beltr\'an~Jim\'enez}},
  \bibinfo{author}{\bibfnamefont{L.}~\bibnamefont{Heisenberg}},
  \bibinfo{author}{\bibfnamefont{T.~S.} \bibnamefont{Koivisto}},
  \bibnamefont{and} \bibinfo{author}{\bibfnamefont{S.}~\bibnamefont{Pekar}},
  \bibinfo{journal}{Phys. Rev. D} \textbf{\bibinfo{volume}{101}},
  \bibinfo{pages}{103507} (\bibinfo{year}{2020}), \eprint{1906.10027}.

\bibitem[{\citenamefont{Nashed}(2010)}]{Nashed:2010ocg}
\bibinfo{author}{\bibfnamefont{G.~G.~L.} \bibnamefont{Nashed}},
  \bibinfo{journal}{Astrophys. Space Sci.} \textbf{\bibinfo{volume}{330}},
  \bibinfo{pages}{173} (\bibinfo{year}{2010}), \eprint{1503.01379}.

\bibitem[{\citenamefont{Alshammari
  et~al.}(2025{\natexlab{b}})\citenamefont{Alshammari, Rizwan
  et~al.}}]{alshammari2025stability}
\bibinfo{author}{\bibfnamefont{M.}~\bibnamefont{Alshammari}},
  \bibinfo{author}{\bibfnamefont{M.}~\bibnamefont{Rizwan}},
  \bibnamefont{et~al.}, \bibinfo{journal}{Int. J. Geom. Methods Mod. Phys.} p.
  \bibinfo{pages}{10.1142/S021988782650009X}
  (\bibinfo{year}{2025}{\natexlab{b}}).

\bibitem[{\citenamefont{El~Hanafy and Nashed}(2016)}]{ElHanafy:2014efn}
\bibinfo{author}{\bibfnamefont{W.}~\bibnamefont{El~Hanafy}} \bibnamefont{and}
  \bibinfo{author}{\bibfnamefont{G.~L.} \bibnamefont{Nashed}},
  \bibinfo{journal}{Astrophys. Space Sci.} \textbf{\bibinfo{volume}{361}},
  \bibinfo{pages}{197} (\bibinfo{year}{2016}), \eprint{1410.2467}.

\bibitem[{\citenamefont{Mandal et~al.}(2020{\natexlab{b}})\citenamefont{Mandal,
  Sahoo, and Santos}}]{Mandal:2020lyq}
\bibinfo{author}{\bibfnamefont{S.}~\bibnamefont{Mandal}},
  \bibinfo{author}{\bibfnamefont{P.~K.} \bibnamefont{Sahoo}}, \bibnamefont{and}
  \bibinfo{author}{\bibfnamefont{J.~R.~L.} \bibnamefont{Santos}},
  \bibinfo{journal}{Phys. Rev. D} \textbf{\bibinfo{volume}{102}},
  \bibinfo{pages}{024057} (\bibinfo{year}{2020}{\natexlab{b}}),
  \eprint{2008.01563}.

\bibitem[{\citenamefont{Barros et~al.}(2020)\citenamefont{Barros, Barreiro,
  Koivisto, and Nunes}}]{Barros:2020bgg}
\bibinfo{author}{\bibfnamefont{B.~J.} \bibnamefont{Barros}},
  \bibinfo{author}{\bibfnamefont{T.}~\bibnamefont{Barreiro}},
  \bibinfo{author}{\bibfnamefont{T.}~\bibnamefont{Koivisto}}, \bibnamefont{and}
  \bibinfo{author}{\bibfnamefont{N.~J.} \bibnamefont{Nunes}},
  \bibinfo{journal}{Phys. Dark Univ.} \textbf{\bibinfo{volume}{30}},
  \bibinfo{pages}{100616} (\bibinfo{year}{2020}), \eprint{2004.07867}.

\bibitem[{\citenamefont{Shirafuji et~al.}(1996)\citenamefont{Shirafuji, Nashed,
  and Kobayashi}}]{Shirafuji:1996im}
\bibinfo{author}{\bibfnamefont{T.}~\bibnamefont{Shirafuji}},
  \bibinfo{author}{\bibfnamefont{G.~G.~L.} \bibnamefont{Nashed}},
  \bibnamefont{and}
  \bibinfo{author}{\bibfnamefont{Y.}~\bibnamefont{Kobayashi}},
  \bibinfo{journal}{Prog. Theor. Phys.} \textbf{\bibinfo{volume}{96}},
  \bibinfo{pages}{933} (\bibinfo{year}{1996}), \eprint{gr-qc/9609060}.

\bibitem[{\citenamefont{Bajardi et~al.}(2020)\citenamefont{Bajardi, Vernieri,
  and Capozziello}}]{Bajardi:2020fxh}
\bibinfo{author}{\bibfnamefont{F.}~\bibnamefont{Bajardi}},
  \bibinfo{author}{\bibfnamefont{D.}~\bibnamefont{Vernieri}}, \bibnamefont{and}
  \bibinfo{author}{\bibfnamefont{S.}~\bibnamefont{Capozziello}},
  \bibinfo{journal}{Eur. Phys. J. Plus} \textbf{\bibinfo{volume}{135}},
  \bibinfo{pages}{912} (\bibinfo{year}{2020}), \eprint{2011.01248}.

\bibitem[{\citenamefont{Gakis et~al.}(2020)\citenamefont{Gakis,
  Kr{\v{s}}{\v{s}}{\'a}k, Levi~Said, and Saridakis}}]{Gakis:2019rdd}
\bibinfo{author}{\bibfnamefont{V.}~\bibnamefont{Gakis}},
  \bibinfo{author}{\bibfnamefont{M.}~\bibnamefont{Kr{\v{s}}{\v{s}}{\'a}k}},
  \bibinfo{author}{\bibfnamefont{J.}~\bibnamefont{Levi~Said}},
  \bibnamefont{and} \bibinfo{author}{\bibfnamefont{E.~N.}
  \bibnamefont{Saridakis}}, \bibinfo{journal}{Phys. Rev. D}
  \textbf{\bibinfo{volume}{101}}, \bibinfo{pages}{064024}
  (\bibinfo{year}{2020}), \eprint{1908.05741}.

\bibitem[{\citenamefont{Yousaf et~al.}(2024)\citenamefont{Yousaf, Bamba,
  Bhatti, and Farwa}}]{yousaf2024quasi}
\bibinfo{author}{\bibfnamefont{Z.}~\bibnamefont{Yousaf}},
  \bibinfo{author}{\bibfnamefont{K.}~\bibnamefont{Bamba}},
  \bibinfo{author}{\bibfnamefont{M.~Z.} \bibnamefont{Bhatti}},
  \bibnamefont{and} \bibinfo{author}{\bibfnamefont{U.}~\bibnamefont{Farwa}},
  \bibinfo{journal}{Int. J. Geom. Methods Mod. Phys.}
  \textbf{\bibinfo{volume}{21}}, \bibinfo{pages}{2430005}
  (\bibinfo{year}{2024}).

\bibitem[{\citenamefont{Yousaf}(2025)}]{yousaf2025viscous}
\bibinfo{author}{\bibfnamefont{Z.}~\bibnamefont{Yousaf}},
  \bibinfo{journal}{Phys. Dark Universe} \textbf{\bibinfo{volume}{48}},
  \bibinfo{pages}{101884} (\bibinfo{year}{2025}).

\bibitem[{\citenamefont{Xu et~al.}(2019)\citenamefont{Xu, Li, Harko, and
  Liang}}]{Xu:2019sbp}
\bibinfo{author}{\bibfnamefont{Y.}~\bibnamefont{Xu}},
  \bibinfo{author}{\bibfnamefont{G.}~\bibnamefont{Li}},
  \bibinfo{author}{\bibfnamefont{T.}~\bibnamefont{Harko}}, \bibnamefont{and}
  \bibinfo{author}{\bibfnamefont{S.-D.} \bibnamefont{Liang}},
  \bibinfo{journal}{Eur. Phys. J. C} \textbf{\bibinfo{volume}{79}},
  \bibinfo{pages}{708} (\bibinfo{year}{2019}), \eprint{1908.04760}.

\bibitem[{\citenamefont{Yousaf et~al.}(2025{\natexlab{c}})\citenamefont{Yousaf,
  Adeel, Rizwan, Mustafa, and Ali}}]{yousaf2025construction}
\bibinfo{author}{\bibfnamefont{Z.}~\bibnamefont{Yousaf}},
  \bibinfo{author}{\bibfnamefont{A.}~\bibnamefont{Adeel}},
  \bibinfo{author}{\bibfnamefont{M.}~\bibnamefont{Rizwan}},
  \bibinfo{author}{\bibfnamefont{G.}~\bibnamefont{Mustafa}}, \bibnamefont{and}
  \bibinfo{author}{\bibfnamefont{A.}~\bibnamefont{Ali}}, \bibinfo{journal}{Int.
  J. Geom. Methods Mod. Phys.} \textbf{\bibinfo{volume}{22}},
  \bibinfo{pages}{2550093} (\bibinfo{year}{2025}{\natexlab{c}}).

\bibitem[{\citenamefont{Xu et~al.}(2020)\citenamefont{Xu, Harko, Shahidi, and
  Liang}}]{Xu:2020yeg}
\bibinfo{author}{\bibfnamefont{Y.}~\bibnamefont{Xu}},
  \bibinfo{author}{\bibfnamefont{T.}~\bibnamefont{Harko}},
  \bibinfo{author}{\bibfnamefont{S.}~\bibnamefont{Shahidi}}, \bibnamefont{and}
  \bibinfo{author}{\bibfnamefont{S.-D.} \bibnamefont{Liang}},
  \bibinfo{journal}{Eur. Phys. J. C} \textbf{\bibinfo{volume}{80}},
  \bibinfo{pages}{449} (\bibinfo{year}{2020}), \eprint{2005.04025}.

\bibitem[{\citenamefont{D'Ambrosio et~al.}(2020)\citenamefont{D'Ambrosio, Garg,
  and Heisenberg}}]{DAmbrosio:2020nev}
\bibinfo{author}{\bibfnamefont{F.}~\bibnamefont{D'Ambrosio}},
  \bibinfo{author}{\bibfnamefont{M.}~\bibnamefont{Garg}}, \bibnamefont{and}
  \bibinfo{author}{\bibfnamefont{L.}~\bibnamefont{Heisenberg}},
  \bibinfo{journal}{Phys. Lett. B} \textbf{\bibinfo{volume}{811}},
  \bibinfo{pages}{135970} (\bibinfo{year}{2020}), \eprint{2004.00888}.

\bibitem[{\citenamefont{Abbott et~al.}(2016)}]{LIGOScientific:2016aoc}
\bibinfo{author}{\bibfnamefont{B.~P.} \bibnamefont{Abbott}}
  \bibnamefont{et~al.} (\bibinfo{collaboration}{LIGO Scientific, Virgo}),
  \bibinfo{journal}{Phys. Rev. Lett.} \textbf{\bibinfo{volume}{116}},
  \bibinfo{pages}{061102} (\bibinfo{year}{2016}), \eprint{1602.03837}.

\bibitem[{\citenamefont{Akiyama et~al.}(2019)}]{EventHorizonTelescope:2019dse}
\bibinfo{author}{\bibfnamefont{K.}~\bibnamefont{Akiyama}} \bibnamefont{et~al.}
  (\bibinfo{collaboration}{Event Horizon Telescope}),
  \bibinfo{journal}{Astrophys. J. Lett.} \textbf{\bibinfo{volume}{875}},
  \bibinfo{pages}{L1} (\bibinfo{year}{2019}), \eprint{1906.11238}.

\bibitem[{\citenamefont{Banados et~al.}(1992)\citenamefont{Banados, Teitelboim,
  and Zanelli}}]{Banados:1992wn}
\bibinfo{author}{\bibfnamefont{M.}~\bibnamefont{Banados}},
  \bibinfo{author}{\bibfnamefont{C.}~\bibnamefont{Teitelboim}},
  \bibnamefont{and} \bibinfo{author}{\bibfnamefont{J.}~\bibnamefont{Zanelli}},
  \bibinfo{journal}{Phys. Rev. Lett.} \textbf{\bibinfo{volume}{69}},
  \bibinfo{pages}{1849} (\bibinfo{year}{1992}), \eprint{hep-th/9204099}.

\bibitem[{\citenamefont{Banados et~al.}(1993)\citenamefont{Banados, Henneaux,
  Teitelboim, and Zanelli}}]{Banados:1992gq}
\bibinfo{author}{\bibfnamefont{M.}~\bibnamefont{Banados}},
  \bibinfo{author}{\bibfnamefont{M.}~\bibnamefont{Henneaux}},
  \bibinfo{author}{\bibfnamefont{C.}~\bibnamefont{Teitelboim}},
  \bibnamefont{and} \bibinfo{author}{\bibfnamefont{J.}~\bibnamefont{Zanelli}},
  \bibinfo{journal}{Phys. Rev. D} \textbf{\bibinfo{volume}{48}},
  \bibinfo{pages}{1506} (\bibinfo{year}{1993}), \bibinfo{note}{[Erratum:
  Phys.Rev.D 88, 069902 (2013)]}, \eprint{gr-qc/9302012}.

\bibitem[{\citenamefont{Maldacena}(1998)}]{Maldacena:1997re}
\bibinfo{author}{\bibfnamefont{J.~M.} \bibnamefont{Maldacena}},
  \bibinfo{journal}{Adv. Theor. Math. Phys.} \textbf{\bibinfo{volume}{2}},
  \bibinfo{pages}{231} (\bibinfo{year}{1998}), \eprint{hep-th/9711200}.

\bibitem[{\citenamefont{Witten}(1998)}]{Witten:1998qj}
\bibinfo{author}{\bibfnamefont{E.}~\bibnamefont{Witten}},
  \bibinfo{journal}{Adv. Theor. Math. Phys.} \textbf{\bibinfo{volume}{2}},
  \bibinfo{pages}{253} (\bibinfo{year}{1998}), \eprint{hep-th/9802150}.

\bibitem[{\citenamefont{Horowitz and Welch}(1993)}]{Horowitz:1993jc}
\bibinfo{author}{\bibfnamefont{G.~T.} \bibnamefont{Horowitz}} \bibnamefont{and}
  \bibinfo{author}{\bibfnamefont{D.~L.} \bibnamefont{Welch}},
  \bibinfo{journal}{Phys. Rev. Lett.} \textbf{\bibinfo{volume}{71}},
  \bibinfo{pages}{328} (\bibinfo{year}{1993}), \eprint{hep-th/9302126}.

\bibitem[{\citenamefont{Horne and Horowitz}(1992)}]{Horne:1991gn}
\bibinfo{author}{\bibfnamefont{J.~H.} \bibnamefont{Horne}} \bibnamefont{and}
  \bibinfo{author}{\bibfnamefont{G.~T.} \bibnamefont{Horowitz}},
  \bibinfo{journal}{Nucl. Phys. B} \textbf{\bibinfo{volume}{368}},
  \bibinfo{pages}{444} (\bibinfo{year}{1992}), \eprint{hep-th/9108001}.

\bibitem[{\citenamefont{Beltr{\'a}n~Jim{\'e}nez
  et~al.}(2018{\natexlab{b}})\citenamefont{Beltr{\'a}n~Jim{\'e}nez, Heisenberg,
  and Koivisto}}]{BeltranJimenez:2017tkd}
\bibinfo{author}{\bibfnamefont{J.}~\bibnamefont{Beltr{\'a}n~Jim{\'e}nez}},
  \bibinfo{author}{\bibfnamefont{L.}~\bibnamefont{Heisenberg}},
  \bibnamefont{and} \bibinfo{author}{\bibfnamefont{T.}~\bibnamefont{Koivisto}},
  \bibinfo{journal}{Phys. Rev. D} \textbf{\bibinfo{volume}{98}},
  \bibinfo{pages}{044048} (\bibinfo{year}{2018}{\natexlab{b}}),
  \eprint{1710.03116}.

\bibitem[{\citenamefont{Awad et~al.}(2017)\citenamefont{Awad, Capozziello, and
  Nashed}}]{Awad:2017tyz}
\bibinfo{author}{\bibfnamefont{A.~M.} \bibnamefont{Awad}},
  \bibinfo{author}{\bibfnamefont{S.}~\bibnamefont{Capozziello}},
  \bibnamefont{and} \bibinfo{author}{\bibfnamefont{G.~G.~L.}
  \bibnamefont{Nashed}}, \bibinfo{journal}{JHEP} \textbf{\bibinfo{volume}{07}},
  \bibinfo{pages}{136} (\bibinfo{year}{2017}), \eprint{1706.01773}.

\bibitem[{\citenamefont{Heisenberg}(2024)}]{Heisenberg:2023lru}
\bibinfo{author}{\bibfnamefont{L.}~\bibnamefont{Heisenberg}},
  \bibinfo{journal}{Phys. Rept.} \textbf{\bibinfo{volume}{1066}},
  \bibinfo{pages}{1} (\bibinfo{year}{2024}), \eprint{2309.15958}.

\bibitem[{\citenamefont{Nashed}(2021)}]{Nashed:2021pkc}
\bibinfo{author}{\bibfnamefont{G.~G.~L.} \bibnamefont{Nashed}},
  \bibinfo{journal}{Astrophys. J.} \textbf{\bibinfo{volume}{919}},
  \bibinfo{pages}{113} (\bibinfo{year}{2021}), \eprint{2108.04060}.

\bibitem[{\citenamefont{Nashed et~al.}(2021)\citenamefont{Nashed, Odintsov, and
  Oikonomou}}]{Nashed:2021gkp}
\bibinfo{author}{\bibfnamefont{G.~G.~L.} \bibnamefont{Nashed}},
  \bibinfo{author}{\bibfnamefont{S.~D.} \bibnamefont{Odintsov}},
  \bibnamefont{and} \bibinfo{author}{\bibfnamefont{V.~K.}
  \bibnamefont{Oikonomou}}, \bibinfo{journal}{Eur. Phys. J. C}
  \textbf{\bibinfo{volume}{81}}, \bibinfo{pages}{528} (\bibinfo{year}{2021}),
  \eprint{2106.13607}.

\bibitem[{\citenamefont{Nashed}(2025)}]{Nashed:2025qeu}
\bibinfo{author}{\bibfnamefont{G.~G.~L.} \bibnamefont{Nashed}},
  \bibinfo{journal}{Phys. Lett. B} \textbf{\bibinfo{volume}{866}},
  \bibinfo{pages}{139566} (\bibinfo{year}{2025}).

\bibitem[{\citenamefont{Zheng and Yang}(2018)}]{Zheng:2018fyn}
\bibinfo{author}{\bibfnamefont{Y.}~\bibnamefont{Zheng}} \bibnamefont{and}
  \bibinfo{author}{\bibfnamefont{R.-J.} \bibnamefont{Yang}},
  \bibinfo{journal}{Eur. Phys. J. C} \textbf{\bibinfo{volume}{78}},
  \bibinfo{pages}{682} (\bibinfo{year}{2018}), \eprint{1806.09858}.

\bibitem[{\citenamefont{Nashed}(2023)}]{Nashed:2023pxd}
\bibinfo{author}{\bibfnamefont{G.~G.~L.} \bibnamefont{Nashed}},
  \bibinfo{journal}{Astrophys. J.} \textbf{\bibinfo{volume}{950}},
  \bibinfo{pages}{129} (\bibinfo{year}{2023}), \eprint{2306.10273}.

\bibitem[{\citenamefont{Kim and Kim}(2012)}]{Kim:2012cma}
\bibinfo{author}{\bibfnamefont{W.}~\bibnamefont{Kim}} \bibnamefont{and}
  \bibinfo{author}{\bibfnamefont{Y.}~\bibnamefont{Kim}},
  \bibinfo{journal}{Phys. Lett.} \textbf{\bibinfo{volume}{B718}},
  \bibinfo{pages}{687} (\bibinfo{year}{2012}), \eprint{1207.5318}.

\bibitem[{\citenamefont{Nashed and Saridakis}(2022)}]{Nashed:2021pah}
\bibinfo{author}{\bibfnamefont{G.~G.~L.} \bibnamefont{Nashed}}
  \bibnamefont{and} \bibinfo{author}{\bibfnamefont{E.~N.}
  \bibnamefont{Saridakis}}, \bibinfo{journal}{JCAP}
  \textbf{\bibinfo{volume}{05}}, \bibinfo{pages}{017} (\bibinfo{year}{2022}),
  \eprint{2111.06359}.

\bibitem[{\citenamefont{York}(1986)}]{York:1986it}
\bibinfo{author}{\bibfnamefont{J.~W.} \bibnamefont{York}, \bibfnamefont{Jr.}},
  \bibinfo{journal}{Phys. Rev. D} \textbf{\bibinfo{volume}{33}},
  \bibinfo{pages}{2092} (\bibinfo{year}{1986}).

\bibitem[{\citenamefont{Li and wang}(2022)}]{Li:2022oup}
\bibinfo{author}{\bibfnamefont{R.}~\bibnamefont{Li}} \bibnamefont{and}
  \bibinfo{author}{\bibfnamefont{J.}~\bibnamefont{wang}},
  \bibinfo{journal}{Phys. Rev. D} \textbf{\bibinfo{volume}{106}},
  \bibinfo{pages}{106015} (\bibinfo{year}{2022}), \eprint{2206.02623}.

\bibitem[{\citenamefont{Wei et~al.}(2022)\citenamefont{Wei, Liu, and
  Mann}}]{Wei:2022dzw}
\bibinfo{author}{\bibfnamefont{S.-W.} \bibnamefont{Wei}},
  \bibinfo{author}{\bibfnamefont{Y.-X.} \bibnamefont{Liu}}, \bibnamefont{and}
  \bibinfo{author}{\bibfnamefont{R.~B.} \bibnamefont{Mann}},
  \bibinfo{journal}{Phys. Rev. Lett.} \textbf{\bibinfo{volume}{129}},
  \bibinfo{pages}{191101} (\bibinfo{year}{2022}), \eprint{2208.01932}.

\bibitem[{\citenamefont{Duan and Ge}(1979)}]{Duan:1979ucg}
\bibinfo{author}{\bibfnamefont{Y.-S.} \bibnamefont{Duan}} \bibnamefont{and}
  \bibinfo{author}{\bibfnamefont{M.-L.} \bibnamefont{Ge}},
  \bibinfo{journal}{Sci. Sin.} \textbf{\bibinfo{volume}{9}}
  (\bibinfo{year}{1979}).

\bibitem[{\citenamefont{Wei and Liu}(2022)}]{Wei:2021vdx}
\bibinfo{author}{\bibfnamefont{S.-W.} \bibnamefont{Wei}} \bibnamefont{and}
  \bibinfo{author}{\bibfnamefont{Y.-X.} \bibnamefont{Liu}},
  \bibinfo{journal}{Phys. Rev. D} \textbf{\bibinfo{volume}{105}},
  \bibinfo{pages}{104003} (\bibinfo{year}{2022}), \eprint{2112.01706}.

\bibitem[{\citenamefont{Schouton}(1951)}]{density}
\bibinfo{author}{\bibfnamefont{J.~A.} \bibnamefont{Schouton}},
  \emph{\bibinfo{title}{{Tensor Analysis for Physicists}}}
  (\bibinfo{publisher}{Claredon,Oxford}, \bibinfo{year}{1951}).

\bibitem[{\citenamefont{Fu et~al.}(2000)\citenamefont{Fu, Duan, and
  Zhang}}]{Fu:2000pb}
\bibinfo{author}{\bibfnamefont{L.-B.} \bibnamefont{Fu}},
  \bibinfo{author}{\bibfnamefont{Y.-S.} \bibnamefont{Duan}}, \bibnamefont{and}
  \bibinfo{author}{\bibfnamefont{H.}~\bibnamefont{Zhang}},
  \bibinfo{journal}{Phys. Rev. D} \textbf{\bibinfo{volume}{61}},
  \bibinfo{pages}{045004} (\bibinfo{year}{2000}), \eprint{hep-th/0112033}.

\bibitem[{\citenamefont{Wald and Zoupas}(2000)}]{Wald:1999wa}
\bibinfo{author}{\bibfnamefont{R.~M.} \bibnamefont{Wald}} \bibnamefont{and}
  \bibinfo{author}{\bibfnamefont{A.}~\bibnamefont{Zoupas}},
  \bibinfo{journal}{Phys. Rev. D} \textbf{\bibinfo{volume}{61}},
  \bibinfo{pages}{084027} (\bibinfo{year}{2000}), \eprint{gr-qc/9911095}.

\bibitem[{\citenamefont{Stuchlik et~al.}(2001)\citenamefont{Stuchlik, Hledik,
  Soltes, and Ostgaard}}]{Stuchlik:2001ap}
\bibinfo{author}{\bibfnamefont{Z.}~\bibnamefont{Stuchlik}},
  \bibinfo{author}{\bibfnamefont{S.}~\bibnamefont{Hledik}},
  \bibinfo{author}{\bibfnamefont{J.}~\bibnamefont{Soltes}}, \bibnamefont{and}
  \bibinfo{author}{\bibfnamefont{E.}~\bibnamefont{Ostgaard}},
  \bibinfo{journal}{Phys. Rev. D} \textbf{\bibinfo{volume}{64}},
  \bibinfo{pages}{044004} (\bibinfo{year}{2001}).

\bibitem[{\citenamefont{Hou et~al.}(2022)\citenamefont{Hou, Zhu, and
  Zhu}}]{Hou:2021bxz}
\bibinfo{author}{\bibfnamefont{S.}~\bibnamefont{Hou}},
  \bibinfo{author}{\bibfnamefont{T.}~\bibnamefont{Zhu}}, \bibnamefont{and}
  \bibinfo{author}{\bibfnamefont{Z.-H.} \bibnamefont{Zhu}},
  \bibinfo{journal}{JCAP} \textbf{\bibinfo{volume}{04}}, \bibinfo{pages}{032}
  (\bibinfo{year}{2022}), \eprint{2112.13049}.

\bibitem[{\citenamefont{Emtsova et~al.}(2023)\citenamefont{Emtsova, Petrov, and
  Toporensky}}]{Emtsova:2022uij}
\bibinfo{author}{\bibfnamefont{E.~D.} \bibnamefont{Emtsova}},
  \bibinfo{author}{\bibfnamefont{A.~N.} \bibnamefont{Petrov}},
  \bibnamefont{and} \bibinfo{author}{\bibfnamefont{A.~V.}
  \bibnamefont{Toporensky}}, \bibinfo{journal}{Eur. Phys. J. C}
  \textbf{\bibinfo{volume}{83}}, \bibinfo{pages}{366} (\bibinfo{year}{2023}),
  \eprint{2212.03755}.

\bibitem[{\citenamefont{Turner and Horne}(2020)}]{Turner:2020gxo}
\bibinfo{author}{\bibfnamefont{G.~E.} \bibnamefont{Turner}} \bibnamefont{and}
  \bibinfo{author}{\bibfnamefont{K.}~\bibnamefont{Horne}},
  \bibinfo{journal}{Class. Quant. Grav.} \textbf{\bibinfo{volume}{37}},
  \bibinfo{pages}{095012} (\bibinfo{year}{2020}).

\bibitem[{\citenamefont{D'Ambrosio et~al.}(2022)\citenamefont{D'Ambrosio, Fell,
  Heisenberg, and Kuhn}}]{DAmbrosio:2021zpm}
\bibinfo{author}{\bibfnamefont{F.}~\bibnamefont{D'Ambrosio}},
  \bibinfo{author}{\bibfnamefont{S.~D.~B.} \bibnamefont{Fell}},
  \bibinfo{author}{\bibfnamefont{L.}~\bibnamefont{Heisenberg}},
  \bibnamefont{and} \bibinfo{author}{\bibfnamefont{S.}~\bibnamefont{Kuhn}},
  \bibinfo{journal}{Phys. Rev. D} \textbf{\bibinfo{volume}{105}},
  \bibinfo{pages}{024042} (\bibinfo{year}{2022}), \eprint{2109.03174}.

\end{thebibliography}

\end{document}